\documentclass[aps,prb,twocolumn,superscriptaddress]{revtex4-2}
\usepackage{amsmath}
\usepackage{bm}
\usepackage{amssymb}
\usepackage{graphicx}
\usepackage{caption}
\usepackage{floatrow}
\usepackage{subcaption}
\usepackage{graphicx,xcolor} 
\usepackage[hidelinks]{hyperref}
\usepackage{array}

\newcommand{\<}{\langle}
\renewcommand{\>}{\rangle}
\newcommand{\cov}{{\rm cov}}
\newcommand{\liq}{{\rm liq}}
\newcommand{\der}{\partial}
\newcommand{\sol}{{\rm sol}}
\newcommand{\cfg}{{\rm cfg}}
\DeclareMathOperator*{\argmax}{argmax}

\DeclareMathOperator*{\sumint}{%
	\mathchoice%
	{\ooalign{$\displaystyle\sum$\cr\hidewidth$\displaystyle\int$\hidewidth\cr}}
	{\ooalign{\raisebox{.14\height}{\scalebox{.7}{$\textstyle\sum$}}\cr\hidewidth$\textstyle\int$\hidewidth\cr}}
	{\ooalign{\raisebox{.2\height}{\scalebox{.6}{$\scriptstyle\sum$}}\cr$\scriptstyle\int$\cr}}
	{\ooalign{\raisebox{.2\height}{\scalebox{.6}{$\scriptstyle\sum$}}\cr$\scriptstyle\int$\cr}}
}

\begin{document}

%\preprint{}

\title{Bayesian inference of composition-dependent phase diagrams}

\author{Timofei Miryashkin}
\email{miriashkin.tn@phystech.edu}
\affiliation{Skolkovo Institute of Science and Technology, Russia}
\affiliation{Moscow Institute of Physics and Technology, Russia}

\author{Olga Klimanova}
\affiliation{Skolkovo Institute of Science and Technology, Russia}
\affiliation{Moscow Institute of Physics and Technology, Russia}

\author{Vladimir Ladygin}
\affiliation{California Institute of Technology, USA}

\author{Alexander Shapeev}
\email{alexander@shapeev.com}
\affiliation{Skolkovo Institute of Science and Technology, Russia}

% \date{\today}
\date{October 31, 2023}

\begin{abstract}
Phase diagrams serve as a highly informative tool for materials design, encapsulating information about the phases that a material can manifest under specific conditions.
In this work, we develop a method in which Bayesian inference is employed to combine thermodynamic data from molecular dynamics (MD), melting point simulations, and phonon calculations, process these data, and yield a temperature-concentration phase diagram.
The employed Bayesian framework yields not only the free energies of different phases as functions of temperature and concentration but also the uncertainties of these free energies originating from statistical errors inherent to finite-length MD trajectories. 
Furthermore, it extrapolates the results of the finite-atom calculations to the infinite-atom limit and facilitates the choice of temperature, chemical potentials, and the number of atoms conducting the next simulation with which will be the most efficient in reducing the uncertainty of the phase diagram.
The developed algorithm was successfully tested on two binary systems, Ge--Si and K--Na, in the full range of concentrations and temperatures.
\end{abstract}

\maketitle

\section{Introduction}

The continued growth of computational power broadens our capabilities across a wide range of research fields.
Computational materials science is an example of such field that has benefited from recent advancements, allowing researchers to explore the phase space of materials with many components, conduct ab initio computations with several hundred atoms, and even simulate complex biological systems. Furthermore, the recent progress in artificial intelligence (AI) has resulted in the integration of AI into various domains of materials science, including ab initio computation, molecular dynamics simulation, and materials informatics. However, the current state of the field presents many open questions that can be approached with AI.

One promising area of research involves applying AI to the construction of phase diagrams \cite{LIU2020113, Arroyave2022}. Phase diagrams play a crucial role in materials design as they serve as maps that reveal which phase is stable under specific temperature, pressure, and concentration. From a theoretical perspective, constructing a phase diagram reduces to determining the free energies of phases.
Once the free energies are obtained, the stable phase or a mixture of phases under given conditions is simply the one that has the minimal total free energy.

The most widely used practical approach is CALPHAD (standing for CALCulation of PHAse
Diagrams), in which free energies are fitted with polynomial-like functions, primarily using experimental data \cite{calphad_book_1998,brief_hist_calphad}. 
The recent advancements of the CALPHAD method aim to incorporate density functional theory (DFT) data in the algorithm, integrate data from various sources, and include uncertainty quantification. 
In the works \cite{Konigsberger_1991,Konigsberger_1995}, a Bayesian approach was utilized in the CALPHAD method to incorporate prior knowledge about the model parameters and to select conditions for the next experiments in an active learning manner.
The authors of \cite{Stan_2003_calphad} proposed a Bayesian method that combines model parameters from different experimental studies and performs uncertainty quantification.
In later studies \cite{ESPEI_2019, Paulson_2019, Otis_2017}, the Markov Chain Monte Carlo statistical learning method was utilized to learn from heterogeneous data sources (experimental, DFT) and to perform uncertainty quantification for the resulting phase diagrams.
 
The thermodynamic integration technique can be employed to accurately reconstruct the free energy from molecular dynamics (MD) or Monte Carlo (MC) simulations \cite{Berend_Smit_MD}.
This technique is based on the fact that, although the free energy is not directly computable in MD, its derivatives with respect to different parameters of the simulations are.
By integrating these derivatives along a path connecting a point of interest and a point where we have information about the free energy values, it is possible to reconstruct the free energy of a phase at every point.
Thermodynamic integration can be applied to a wide range of systems, including the Lennard-Jones system \cite{lj_phad}, water \cite{deep_water_potential}, nitrogen \cite{Kruglov_nitrogen}, oxides \cite{MgO_CaO_phad}, and even high-entropy alloys \cite{grabowski2019-free-energy-hea}.

% Along with thermodynamic integration, there are other methods for calculating phase diagrams.
% In the coexistence simulation technique the phase transition temperature is found as the temperature at which two investigated phases coexist in equilibrium; this approach was utilized to establish the pressure-temperature (p-T) phase diagram of silicon \cite{silicon_phad_2018_bartok} and temperature-concentration (T-c) binary diagram of Ni-Mo \cite{Ni-Mo_phad_2018_Li}. 
% The interface pinning method \cite{interface_pinning_2013} may be viewed as the modification of the coexistence method in which the free energy difference is determined as a result of the coexistence simulation. The interface pinning method was employed to construct the Ga-As binary phase diagram \cite{Ga-As_phad_2021}. 
% Another approach is the nested sampling algorithm, which determines the phase transition temperature by evaluating the discrepancy in the first derivatives of the free energy \cite{nested_sampling_2016, Rosenbrock2021}; the algorithm was recently applied to construct the melting curves of the Ag-Pd system \cite{Rosenbrock2021}. 
% High-throughput approach developed in [] uses the mean field model (accounting only for configurational entropy) to estimates the transition temperatures of equiatomic multi-component solid solution.   

Along with thermodynamic integration, there are other methods for calculating phase diagrams.
In the coexistence simulation technique the phase transition temperature is found as the temperature at which two investigated phases coexist in equilibrium; this approach was utilized to establish the pressure-temperature (p-T) phase diagram of silicon \cite{silicon_phad_2018_bartok} and temperature-concentration (T-c) binary diagram of Ni-Mo \cite{Ni-Mo_phad_2018_Li}. 
The interface pinning method \cite{interface_pinning_2013, Ga-As_phad_2021} may be viewed as the modification of the coexistence method in which the free energy difference is determined as a result of the coexistence simulation. 
Another approach is the nested sampling algorithm, which determines the phase transition temperature by evaluating the discrepancy in the first derivatives of the free energy \cite{nested_sampling_2016, Rosenbrock2021}. 
A mean field model, accounting only for configurational entropy, was proposed in \cite{Lederer_high_throughput_approach} and applied to estimate the transition temperatures of binary and equiatomic multicomponent solid solutions in a high-throughput manner.

The current study introduces a Bayesian learning algorithm for constructing T-c phase diagrams, which can be viewed as an extension of the approach presented in \cite{ladygin2021-phad}, previously applied to p-T diagrams.
Our algorithm is based on Gaussian process regression, which not only reconstructs the free energy from the MD/MC data, but also propagates the statistical uncertainty in the data to the uncertainty of the free energy and, subsequently, to the phase boundaries.
Our algorithm accepts various data as input including melting points, ensemble-averaged potential energy and concentration, and results of phonon calculations.
Our method is further equipped with an active learning algorithm that can suggest new points for MD/MC calculations to reduce the uncertainty in the phase diagram prediction in the most effective way.

We validate our algorithm on the Ge--Si and K--Na binary systems.
We start with the relatively simple Ge--Si phase diagram featuring the miscibility gap at low temperatures which, at higher temperatures, turns to the solid solution on the entire composition range which melts upon further temperature increase.
The K--Na system poses a more challenging test for our algorithm due to its non-trivial features: it includes two distinct bcc-Na and bcc-K phases, an intermetallic phase, several regions in which the solid and liquid phases coexist, and a eutectic point at which the Na$_2$K and bcc-Na solid phases simultaneously solidify.
%In the current algorithm implementation, we use only thermodynamic data; however, experimental data can also be seamlessly added to increase the accuracy of the resulting phase diagram.

The article is organized as follows. In Section \ref{sec:theory} we outline the methodology and provide the implementation details. Section \ref{sec:results} 
presents the results of applying our algorithm to the Ge--Si and K--Na binary systems. Summary and concluding remarks are given in Section \ref{sec:conclusion}.

\section{Theory}
\label{sec:theory}

\subsection{Definition of free energy}

We consider a binary system with two types of atoms.
The number of atoms is denoted by $N$, the positions by $x_i$ ($i=1,\ldots,N$), and atomic types by $\sigma_i \in \{1,2\}$.
The volume occupied by atoms is denoted as $\hat{V}$, and we represent the configuration as
\[
\cfg := (\hat{V},x_1,\ldots,x_N,\sigma_1,\ldots,\sigma_N).
\]
Here and in what follows the sign ``:='' means ``equal by definition''.
The (potential) energy of interatomic interaction of the configuration $\cfg$ is denoted as $\hat{E}(\cfg)$ and the energy-per-atom is $E(\cfg) = \hat{E}(\cfg)/N$.
Here and in what follows by $\hat{\vphantom{t}\bullet}$ we denote extensive quantities.
The number of type-$2$ atoms in a configuration $\cfg$ is denoted by $ \hat{\chi} (\cfg)$, so that their concentration is $\chi(\cfg) = \hat{\chi}(\cfg)/N$.
The concentration of type-1 atoms is hence $1 - \chi$.

Our goal is to reconstruct the free energy $F$ as a function of temperature and composition.
To that end, we will use the semi-grand-canonical MC+MD simulations sampling the $\mu$pT ensemble.
%For simplicity, we will call the grand-canonical MC+MD simulation simply a grand-canonical simulation or GCS.
The semi-grand-canonical $\mu$pT simulation (referred to as simply the simulation in what follows) is driven by the chemical potentials of the species. Without loss of generality, we always assume that the chemical potential of type-1 atoms is $\mu_1 = 0$ and for simplicity we denote $\mu := \mu_2$ as the chemical potential of the type-2 species.
The $\mu$pT simulation effectively samples the system's energy with a probability of configuration $\cfg$ proportional to ${\rm exp}(-\beta (\hat{E}(\cfg) - \mu\cdot \hat{\chi}(\cfg)))$, where $\beta=1/T$, $T$ is the temperature, and we work in the units in which the Boltzmann constant is $k_{\rm B} = 1$.

As an intermediate quantity, we define the semi-grand potential $\Phi = \Phi(T,\mu)$ as
\begin{equation}\label{eq:Phi}
	\hat{\Phi} := - \beta^{-1} \log \hat{Z},
\end{equation}
where
\begin{equation}\label{eq:partition-function}
\hat{Z} := \sumint_{\cfg} \exp\big(-\beta (\hat{E}(\cfg) - \mu\cdot \hat{\chi}(\cfg) + p\hat{V}) \big).
\end{equation}
Here $\sumint_{\cfg}$ expands to
\[
\sum_{\sigma_1}\cdots \sum_{\sigma_N} \ \
\int_0^{\infty} {\rm d}\hat{V} \ \ 
\int_{x_1\in\hat{V}}{\rm d}x_1\cdots\int_{x_N\in \hat{V}}{\rm d}x_N.
\]
In this work, we are interested in small, negligible $p$, therefore in what follows we simply assume that $p=0$.

From the simulations it is easy to compute averages of microscopic quantities.
We denote by $f(\cfg)$ any function of a configuration and define its ensemble average as
\[
\<f\> = \<f\>_{T,\mu} = Z^{-1} \sumint_{\cfg} f(\cfg) \exp\big(-\beta (\hat{E}(\cfg) - \mu\cdot \hat{\chi}(\cfg)) \big).
\]
The first quantity we would be interested in is the composition
$c := \<\chi\>$.
We denote the free energy, as a function of composition, by $\hat{G}(T,c)$, which may be referred to as the Gibbs free energy.
It is related, by its definition, to $\Phi$ as
\begin{equation}\label{eq:F}
\hat{G}(T,\<\chi\>_{T,\mu}) := \hat{\Phi}(T,\mu) + \mu\cdot \<\hat{\chi}\>_{T,\mu}.
\end{equation}
Sometimes it may be important to explicitly keep track of the dependence of the free energy on the number of atoms $N$, in which case it it will be denoted as $\hat{G}(T, c, N)$ (and similarly for other quantities).

\subsection{Free energy asymptotics}
\label{subsec:f_ref}

We build a physically inspired AI algorithm that incorporates our knowledge of the asymptotics of the free energies. To that end, we find it convenient to work with the quantity $S(T, c, N)$ defined as 
\begin{align}
    G(T, c, N) = G_{\text{ref}}(T, c, N) - T S(T, c, N),
    \label{eq:S_def}
\end{align}
where $G_{\text{ref}}(T, c, N)$ is a reference free energy.
If $G_{\text{ref}}(T, c, N)$ is chosen as $G(0, c, N)$ then $S(T, c, N)$ is the conventional entropy.
If $G_{\text{ref}}(T, c, N)$ is chosen to be the free energy of an ideal gas, then the resulting $S(T, c, N)$ would typically be called the excess entropy.
By analogy, we refer to $S(T, c, N)$ as \emph{entropy}, although our $G_{\text{ref}}(T, c, N)$ will be somewhat different from the common ones and will be chosen to facilitate the fitting of $S(T, c, N)$.
% . Our algorithm incorporates the knowledge of the free energy asymptotic behavior in $G_{\text{ref}}(T, c, N)$, while $S(T, c, N)$ is learned from its derivatives.
In the remainder of this subsection, we show how we select $G_{\text{ref}}(T, c, N)$ for the crystalline, intermetallic, and liquid phases.

For the crystalline phases (bcc/fcc/diamond), the free energy as $T \rightarrow 0$ has the form 
\begin{align}\label{eq:f_cryst}
\begin{split}
	G^{\rm cryst} (T, c, N)
	&= 
	E_0
    + T c \log(c)
    + T (1 - c) \log(1 - c) \\
	&- T \log(N) + T - {\textstyle \frac32} T \log(2\pi T) 
    \\&
    + {\textstyle \frac12} T N^{-1} \log\det \hat{H}_0(N)
	\\&
	 + O\big(T\big)o(1) + O(T^2)
	 ,
\end{split}
\end{align}
where $E_0$ is the energy of the ground state structure, $H_0$ is the energy Hessian at the ground state and $o(1)$ denotes a vanishing as $N\to\infty$ term. 
The derivation of \eqref{eq:f_cryst} is given in Appendix \ref{app:free_en_asym}.
Note that $ \log\det \hat{H}_0(N)$ scales as $O(N)$, therefore $N^{-1} \log\det \hat{H}_0(N)$ is a nonvanishing intensive quantity.
The formula above explicitly accounts for the vibrational and configurational contributions to the free energy.
We select the following reference free energy for the crystalline phase:
\begin{align*}
\begin{split}
G^{\rm cryst}_{\rm ref}(T, c, N)
&:= 
E_0(c)
+ T c \log(c)
+ T (1-c) \log(1-c) \\
&- T \log(N) + T - {\textstyle \frac32} T \log(2\pi T),
\end{split}
\end{align*}
where $E_0(c)$ is the linear interpolation between the energies of monoatomic ground-state structures.

% We next obtain the asymptotic free energy for the equiatomic intermetallic alloy.
% In the formula below we assume that the intermetallic structure admits two types of substitutional defects, the first one is a substitution of type-2 atom by a type-1 atom and has the formation energy of $E_1$, while the second one substitutes the type-1 with type-2 and has the formation energy of $E_2$.
% The asymptotics for the free energy is 
% \begin{align*}
% \begin{split}
% G^{\rm im}(T, c, N) &= E_0  
% - T \log \big( 1 + e^{- \beta \frac{ (E_1 + E_2)}{2}} \big)
% + \frac{(E_1 - E_2)}{2} \left(c - \frac12\right) \\
% &+ T \bigg( 1 + \cosh \bigg( \beta \frac{ (E_1 + E_2) }{2} \bigg) \bigg) \left(c - \frac12\right)^2 \\
% &- T \log(N) + T - {\textstyle \frac32} T \log(2\pi T) 
% + {\textstyle \frac12} T N^{-1} \log\det \hat{H}_0 \\
% &+ O\left( T^2 + N^{-1/2} + \left(c-\frac12\right)^3\right).
% \end{split}
% \end{align*}
% and its derivation is given in in Appendix \ref{appd:asym_im}.
% The asymptotic free energy for the arbitrary intermetallic alloy may be derived following steps in Appendix \ref{appd:asym_im}. The reference free energy of intermetallic alloy is chosen by analogy to solid.

For the liquid phase, we select the reference free energy that accounts only for the configurational contribution 
\begin{align*}
\begin{split}
G^{\rm liq}_{\rm ref}(T, c, N)
&:= 
T c \log(c) + T (1-c) \log(1-c) 
\\&- T \log(N).
\end{split}
\end{align*}

As with the thermodynamic integration method, the $\mu$pT simulations provide information on the derivative of $S(T,c, N)$ with respect to $T$ and $c$.
This information is sufficient to reconstruct $S(T,c,N)$ for each phase up to an additive constant, however, to accurately find the phase boundaries we need to also find this constant.
For the solid phases (crystalline and intermetallic), we do this from the harmonic limit
\begin{align}\label{eq:ref_solid}
\begin{split}
S(0,c,N) &=
\lim_{T\to0} T^{-1} (F_{\rm ref}(T,c,N) - G(T,c,N))
\\&=
-{\textstyle \frac12} N^{-1} \log\det \hat{H}_0(N).
\end{split}
\end{align}
For liquid we do it from the melting point data, which we obtain from NpT coexistence simulations, as described in Section \ref{subsec:melt_points}.

\subsection{Thermodynamic integration}
\label{subsec:thermo_integ}

The derivatives of the free energy have the form
% \begin{align}
%     \begin{split}
%         \frac{\partial G(T, \<\chi\>)}{\partial \<\chi_i\>} &= \mu_i \\ \hfil
%         \frac{\partial (\beta G(T, \<\chi\>))}{\partial T} &= -\frac{\< E \>}{T^2},    
%     \end{split} 
% \end{align}
\begin{align}
    \frac{\partial G(T, \<\chi\>)}{\partial \<\chi\>} &= \mu,
    \label{eq:dF_dc}
    \\
    \frac{\partial (\beta G(T, \<\chi\>))}{\partial T} &= -\frac{\< E \>}{T^2},
    \label{eq:dF_dT}
\end{align}
as we show in Appendix \ref{appd:entrop_deriv}.
Using the definition of the entropy \eqref{eq:S_def} along with formulas \eqref{eq:dF_dc} and \eqref{eq:dF_dT}, we obtain the following expressions for the derivatives of the entropy
\begin{align*}
	\frac{\partial S}{\partial T} &= \frac{\<E\>}{T^2} + \frac{\partial (\beta G^{\rm ref})}{\partial T},
	% \label{eq:dS_dT}
	\\ \notag
	\frac{\partial S}{\partial c} &= - \beta \mu + \frac{\partial (\beta G^{\rm ref})}{\partial c}.
	% \label{eq:dS_dc}
\end{align*}

In our Gaussian process regression-based algorithm we are able to incorporate the uncertainty, in the sense of statistical error of averaging over finite MD trajectories. We distinguish between the mathematical expectation (``true mean'') of concentration and energy, $\< \chi \>$ and $\<E\>$, and the trajectory-averaged mean, $\overline{\chi}$ and $\overline{E}$, which can be treated as random variable. We note that $\big\<\overline{E}\big\> = \<E\>$---in other words, the expected sample mean of $E$ is the true mean of $E$, although, e.g., ${\rm cov}\big(\overline{E}, \overline{E}\big) \ne {\rm cov}(E, E)$---i.e., the variance of the sample mean is different, and should be much smaller than ${\rm cov}(E, E)$.
We derive in Appendix \ref{appd:error_input_data} the following formulas for the uncertainty of determining the derivatives of $S$,
\begin{align*}\notag
	\Delta\left( \frac{\partial S}{\partial T} \right)
	&=
	\Bigg(\frac{{\rm cov}\big(\overline{E}, \overline{E}\big)}{T^4}
	+ \frac{2}{T^2} \dfrac{\partial^2 (\beta G)}{\partial T \partial c} {\rm cov}\big(\overline{E}, \overline{c}\big)
	\\&\,\hphantom{=\Bigg(}\,
	+ \left( \frac{\partial^2 (\beta G)}{\partial T \partial c}\right)^2\Bigg)^{1/2}
	\big({\rm cov}(\overline{c}, \overline{c})\big)^{1/2},
	% \label{eq:err_ds_dt}
	\\ \notag
	\Delta \left( \frac{\partial S}{\partial c} \right)
	&=
	\left| \dfrac{\partial^2 G}{\partial c^2} \right| \, \big({\rm cov}(\overline{c}, \overline{c})\big)^{1/2}.
	% \label{eq:err_ds_dc}
\end{align*}
We will make these formulas more rigorous in the following subsections where we treat $S$ as a random variable distributed according to a Gaussian process.

%We introduce $\mathcal{E} = \{\overline{E}_1, \ldots, \overline{E}_n\}$, where $\overline{E}_i$ is the sample means over i-th trajectory, while the $\mathcal{E}$ can be viewed as the random variable with i.i.d. samples $\overline{E}_i$. By analogy, we introduce random variable $\mathcal{C} = \{\overline{\chi}_1, \ldots, \overline{\chi}_n\}$ with i.i.d. samples $\overline{\chi}_i$. Our algorithm based on the Gaussian process incorporates the error of the entropy derivatives, which have the following form (appendix \ref{appd:error_input_data})

\subsection{Gaussian process regression}
\label{subsec:GPR}

We use Gaussian process regression \cite{rasmussen2006-book-gaussian} to reconstruct the entropy of each phase from the input data while accounting for their uncertainties.
The data in simulations is close to the normal distribution, therefore the Gaussian process is a natural approach to reconstruct the entropy.

In the Gaussian process, we model the covariance of the entropy between two points $(T_1, c_1, N_1)$ and $(T_2, c_2, N_2)$ using the kernel
\[
	\begin{array}{r}
k[(T_1, c_1, N_1), (T_2, c_2, N_2)]
\hspace{12em}\mathstrut
\\
:= {\rm Cov}[S(T_1, c_1, N_1), S(T_2, c_2, N_2)].
\end{array}
\]
We assume that the entropy is a smooth function of $T$, $c$, and $N^{-1}$, therefore we use the square exponential function to capture the covariance between two points. For the crystalline phase, the kernel has the form
\begin{align}\label{eq:cryst_ker}
\begin{split}
    k_{\rm cryst}&:=  \theta_0^2 + \theta_f^2  
    \exp \left( -\frac{(T_1 - T_2)^2}{2 \theta_T^2}\right) 
    \exp \left( -\frac{(c_1 - c_2)^2}{2 \theta_c^2} \right) 
    \\&\phantom{:=\mathstrut} 
    \cdot \exp  \left( -\left(\frac{1}{N_1} - \frac{1}{N_2} \right)^2 \frac{\theta_N^2}{2} \right),
\end{split}
\end{align}
where we also include the $\theta_0$ hyperparameter into the kernel to capture the constant shift in the data that may be of a different scale than $\theta_f$.
It also should be noted that the kernel is constructed in such a way that it allows for substituting $N=\infty$ (or, to be more precise, taking a limit $N\to\infty$).
Thanks to this, it will be possible to conduct simulations for different numbers of atoms $N$ and extrapolate the result to an infinite number of atoms.

% For the liquid it is important to explicitly take into account that
% the energy is defined up to two additive constants (an arbitrary energy shift for each of the species).  Therefore we allow the entropy to have a linear in $c$ shift, $S \sim \theta_1 c/T + \theta_2 (1-c)/T$, which features as the last two terms in our kernel for the liquid phase:

We note that for a unary system, the potential energy of a configuration $\hat{E}(\cfg)$ can be arbitrarily shifted by a constant, while for a binary system the energy can be shifted by $\theta_1 \hat{\chi}(\cfg) + \theta_2 (1-\hat{\chi}(\cfg))$, where $\theta_1$ and $\theta_2$ are the constant energies of each of the species. For liquid it is important to explicitly take this into account treating $\theta_1$ and $\theta_2$ as hyperparameters. We therefore assume that the entropy has the linear in $c$ shift, $S \sim \theta_1 c/T + \theta_2 (1-c)/T$, which features as the last two terms in our kernel for the liquid phase:
\begin{align*}
    k_{\rm liq} &:=  \theta_0^2 + \theta_f^2  
    \exp \left( -\frac{(T_1 - T_2)^2}{2 \theta_T^2}\right) 
    \exp \left( -\frac{(c_1 - c_2)^2}{2 \theta_c^2} \right)  
\\ &\mathstrut\phantom{:=\mathstrut}\mathstrut
    \cdot \exp  \left( -\left(\frac{1}{N_1} - \frac{1}{N_2} \right)^2 \frac{\theta_N^2}{2} \right)
\\ &\mathstrut\phantom{:=}\mathstrut
	+ \theta_{1}^2 \frac{c_1 c_2}{T_1 T_2} 
    + \theta_{2}^2 \frac{(1-c_1) (1-c_2)}{T_1 T_2}.
\end{align*}

The advantageous characteristic of a Gaussian process is that any linear functional of a Gaussian process is also Gaussian distributed. This property is useful in our work as the input data has the form of the entropy derivatives. For instance, the covariance between the entropy derivative and the entropy itself is given by the following kernel
\[
{\rm Cov}\left[\frac{\partial }{\partial T_1} S(T_1, c_1), S(T_2, c_2)\right] = \frac{\partial }{\partial T_1} k[(T_1, c_1), (T_2, c_2)].
\]
In the general case, a data point may be viewed as a linear functional $X$ on $S$, for example, $\< S | X_1\> = S(T, c) $, $\< S | X_2\> = \frac{\partial}{\partial T} S(T, c) $.
We thus extend the definition of the kernel to arbitrary functionals $X_1$ and $X_2$:
\[
k(X_1, X_2) := {\rm Cov}[\< S| X_1 \>, \< S| X_2 \>].
\]

We now discuss the Gaussian process regression algorithm in which we use the kernels defined above.
We perform calculations in the data point (functional) $X_i$ for which we obtain the target value $Y_i$ with the Gaussian noise $\Delta Y_i$, and hence require that the input point has the Gaussian distribution:
\[
\< S | X_i\> \sim \mathcal{N} \big( Y_i, (\Delta Y_i)^2 \big).    
\]
Thus, the input data to the regression algorithm has the form of tuples $(X_i, Y_i, \Delta Y_i)$. We next want to make a prediction $Y_* = \< S|X_*\>$ in the point $X_*$ which we may not have previously observed. 
To that end, we form a joint Gaussian distribution for the input points and the point for prediction.
It has zero mean and the convariance matrix given by
%\begin{align*}
%    \left[\begin{array}{c}
%    \boldsymbol{Y} \\
%    Y_*
%    \end{array}\right] &\sim \mathcal{N}\left(\bigg(\begin{array}{l}
%    \mathbf{0} \\
%    0
%    \end{array}\bigg),
%    \left(\begin{array}{cc}
%    K(\boldsymbol{X}, \boldsymbol{X})+\operatorname{diag}(\boldsymbol{\Delta} \boldsymbol{Y}^2) & K\left(\boldsymbol{X}, X_*\right) \\
%    K\left(X_*, \boldsymbol{X}\right) & K\left(X_*, X_*\right)
%    \end{array}\right)\right),
%\end{align*}
\[
	{\rm cov}
	\begin{pmatrix}
		\boldsymbol{Y} \\
		Y_*
	\end{pmatrix}
=
	\begin{pmatrix}
		K(\boldsymbol{X}, \boldsymbol{X})+\operatorname{diag}(\boldsymbol{\Delta} \boldsymbol{Y}^2) & K\left(\boldsymbol{X}, X_*\right) \\
		K\left(X_*, \boldsymbol{X}\right) & K\left(X_*, X_*\right)
	\end{pmatrix},
\]
where ${\bm X}$ and ${\bm Y}$ are the vectors composed of $X_i$ and $Y_i$, $\operatorname{diag}(\boldsymbol{\Delta Y}^2)$ is the matrix with $(\Delta Y_i)^2$ on the diagonal, and  $K(\bm{X}, \bm{X})$ is the matrix composed of $k(X_i, X_j)$. 
From this, we follow \cite{bishop2006pattern} and obtain that $Y_*$ is again Gaussian with the mean and variance given by
\begin{align} \notag
    \mathbb{E}[Y_*] &= K(X_*, {\bm X}) K_y^{-1} {\bm Y}, \qquad{\text{and}}
\\ \label{eq:var_gp}
    \mathbb{V}[Y_*] &= K(X_*, X_*) - K(X_*, {\bm X}) K_y^{-1} K({\bm X}, X_*), 
\end{align}
where $K_y = K({\bm X}, {\bm X}) + {\rm diag} ({\bm{\Delta Y^2} })$.

Furthermore, we can estimate the variance of a nonlinear functional $\mathcal{F}(S)$ of the Gaussian process. The original functional is linearized by taking the first term of the Taylor series expansion around the mean of the Gaussian process  
\[
\mathcal{F}(S) \approx \mathcal{F}(\overline{S}) + \< S - \overline{S} | \nabla_S \mathcal{F} (\overline{S}) \>, 
\]
where $\overline{S}$ is the mean predicted entropy, $\nabla_S \mathcal{F} (\overline{S})$ is the Jacobian of $\nabla_S \mathcal{F}$ evaluated at $\overline{S}$. The variance of the linearized functional has the form similar to \eqref{eq:var_gp} 
\begin{align}
    \mathbb{V}[\mathcal{F}] \approx \mathbb{V}[{\bm J}] = K({\bm J}, {\bm J}) - K({\bm J}, {\bm X}) K_y^{-1} K({\bm X}, {\bm J}),
    \label{eq:var_j}
\end{align}
where for simplicity we denote ${\bm J} : = \nabla_S \mathcal{F} (\overline{S})$. 

Gaussian process regression is a non-parametric machine learning algorithm that depends only on hyperparameters, for example ${\bm \theta} = (\theta_0, \theta_f, \theta_T, \theta_c, \theta_N)$ in equation \eqref{eq:cryst_ker}. Hyperparameters are optimized by maximizing the marginal likelihood $p({\bm Y} | {\bm X}, {\bm \theta})$, which gives the probability of observing the targets ${\bm Y}$ given the feature matrix $\bm{X}$ and hyperparameters ${\bm \theta}$. The logarithm of the marginal likelihood is expressed as
\begin{align*}
    \log p({\bm Y}| {\bm X}, {\bm \theta}) &= - {\textstyle \frac12} {\bm Y}^{T} K_y^{-1} {\bm Y} 
    - {\textstyle \frac12} \log |K_y| 
    - {\textstyle \frac{n}{2}} \log(2\pi),
\end{align*}
where $n$ is the number of the input points.

\subsection{Melting points}
\label{subsec:melt_points}

\begin{figure*}
	\includegraphics[width=15cm]{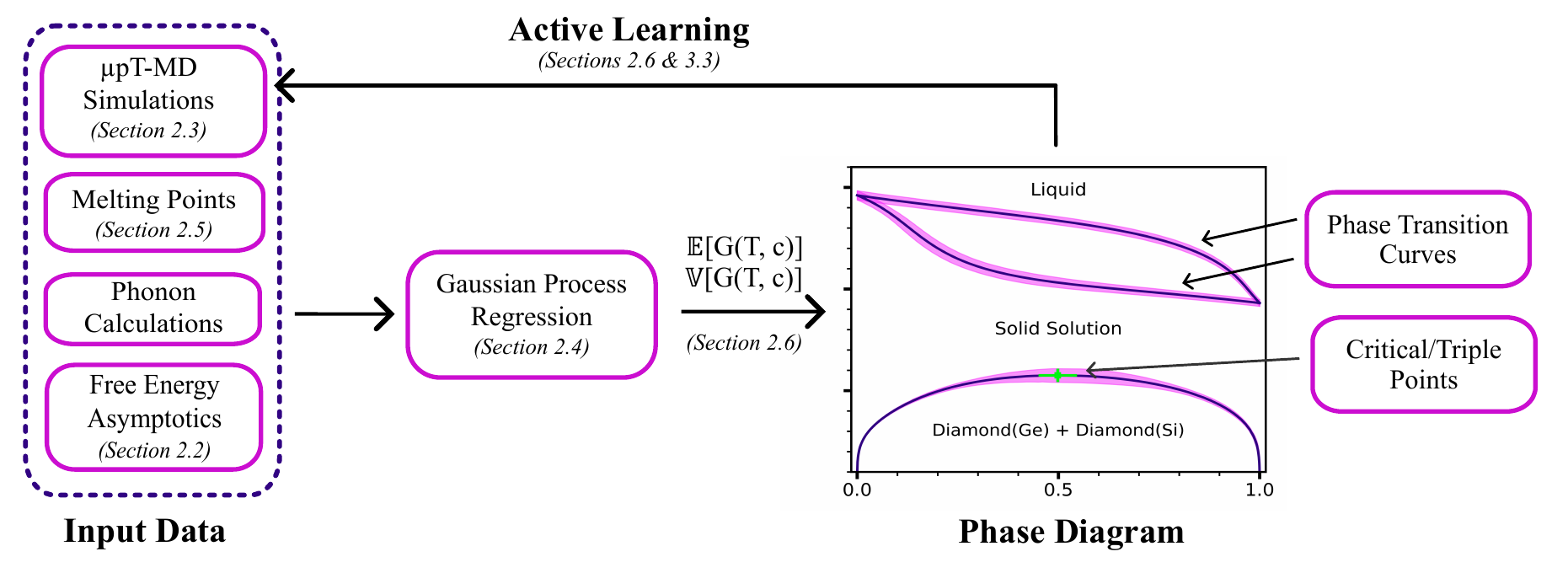}
    \caption{Schematic illustration of our algorithm. Thermodynamic data of different natures (MD simulations, melting points, phonon computations) along with the asymptotics of the free energies is fed as input data to the Gaussian process regression. We next predict the free energies with their uncertainties to construct the phase diagram. Additionally, following the optimal uncertainty reduction strategy, the algorithm suggests the points on the phase diagram where the next MD simulations should be conducted.}
    \label{fig:algo_scheme}
\end{figure*}

Thanks to the Bayesian approach, our algorithm incorporates the data on melting points along with their corresponding uncertainties. We implement the melting point calculation following the method presented in \cite{klimanova_melt_points}.

The approach consists of running a number of NpT-MD simulations for unary compounds ($c_{\rm m} = 0$ or $c_{\rm m} = 1$) starting from an atomistic system consisting of liquid and solid phases in roughly equal proportions.     
Each simulation is run until the simulation box contains either all-solid phase or all-liquid phase (which we refer to as the ``solid outcome'' or ``liquid outcome'').
The data from these simulations is collected into a table with the values of $N$, $T$, and the number of solid and liquid outcomes.
These data are then processed by a comprehensive nonlinear Bayesian optimization algorithm described in detail in \cite{klimanova_melt_points}, which results in the prediction of the melting point $T_{\rm m}$ and its associated uncertainty $\Delta T_{\rm m}$ in the limit of an infinite number of atoms ($N\to\infty$).

The melting points along with their corresponding uncertainties are then added into our algorithm as the linear functional of the Gaussian process
\[
G_2 (T_{\rm m}, c_{\rm m}, N) - G_1 (T_{\rm m}, c_{\rm m}, N) = 0,
\] 
with the uncertainty of the functional given by
\[
\Delta\left( G_2(T_{\rm m}, c_{\rm m}, N) - G_1(T_{\rm m}, c_{\rm m}, N) \right) = \left| \frac{\partial G_2}{\partial T} - \frac{\partial G_1}{\partial T} \right| \Delta T_{\rm m}.
\]
This effectively propagates the uncertainty in the melting point predictions to the phase diagram.

\subsection{Bayesian learning algorithm: bringing all data together}

\begin{figure}[h!]
	\centering
	\includegraphics[width=8cm]{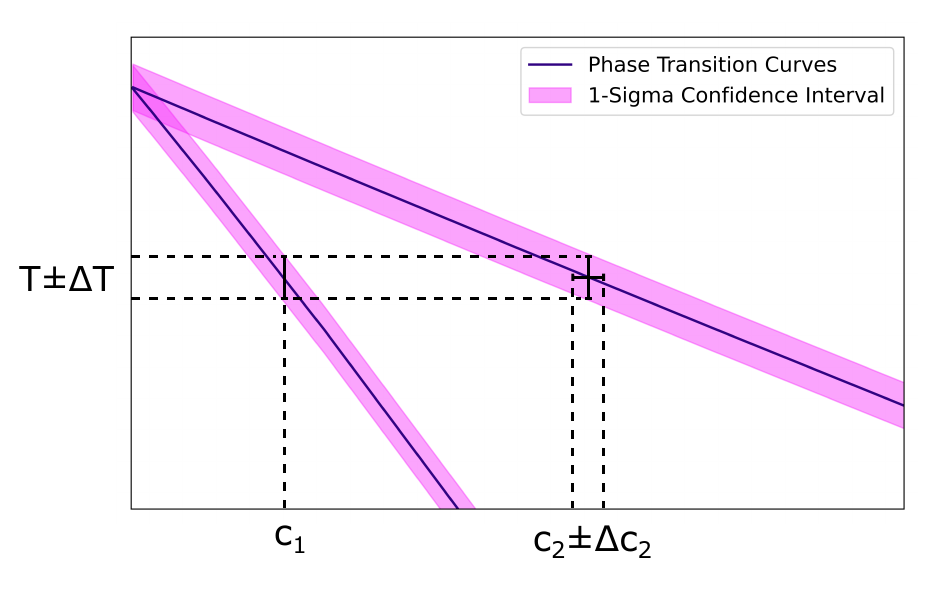}
    \caption{Illustration of how the phase boundaries are determined with their confidence intervals: we fix $c_1$ and find $(T, c_2)$ by solving \eqref{eq:sys_eq_1}--\eqref{eq:sys_eq_2}, then the uncertainty of $(T, c_2)$ is estimated by \eqref{eq:var_j}.}
    \label{fig:sys_eq}
\end{figure}

Here, we present a general scheme of our algorithm, which is illustrated in Figure \ref{fig:algo_scheme}. 
Our algorithm accepts input data of different nature with the corresponding uncertainties: derivatives of the entropy obtained from $\mu$pT simulations (Section \ref{subsec:thermo_integ}), melting points (Section \ref{subsec:melt_points}), and phonon calculations (which is precisely the calculation of the energy Hessian in \eqref{eq:ref_solid}), along with the free energies asymptotics (Section \ref{subsec:f_ref}).
Hyperparameters of our algorithm are optimized with respect to the input data as described in Section \ref{subsec:GPR}; after training the algorithm reconstructs the free energy and its derivatives (with uncertainties), which we utilize further to predict the phase diagram.

We will now describe the inference of the phase transition curves, while inference for triple and critical points is done by analogy. We determine the temperature and concentrations at which the phase transition takes place between two phases.
We let the temperature of the transition be $T$ and the corresponding concentrations be $c_1$ and $c_2$, respectively.
Denoting the free energies of the two phases as $G_1$ and $G_2$, the transition curve can be determined by solving the system of equations:
\begin{align}\label{eq:sys_eq_1}
	\mathcal{K}_1 & := \frac{ \partial (G_2/T)}{\partial c} (T, c_2) - \frac{ \partial (G_1/T)}{\partial c} (T, c_1) = 0,
	\\ \notag
	\mathcal{K}_2 & := \frac{1}{T} \bigg( G_2(T, c_2) - G_1(T, c_1) - c_2 \frac{ \partial G_2}{\partial c}(T, c_2)
    \\&\phantom{:=\frac{1}{T}\bigg(\mathstrut} \label{eq:sys_eq_2}
    + c_1 \frac{\partial G_1}{\partial c} (T, c_1) \bigg) = 0,
\end{align}
where we denoted the left-hand size of these equations as $\mathcal{K}_1$ and $\mathcal{K}_2$.
To resolve the transition curve, we can, for example, fix the concentration $c_1$ and solve the above equations for the phase transition point ${\bm p} = (T, c_2)^T$ as shown in Figure \ref{fig:sys_eq}. Our algorithm estimates the uncertainty of the solution by utilizing \eqref{eq:var_j}, where $\bm{J}$ is expressed as  
\[
\bm{J} = \nabla_S {\bm p} = - (\nabla_{\bm p} \mathcal{K} )^{-1} (\nabla_S \mathcal{K}),
\]
where $\mathcal{K} = (\mathcal{K}_1, \mathcal{K}_2)^T$ is the joint left-hand side of \eqref{eq:sys_eq_1} and \eqref{eq:sys_eq_2}.
The covariance of ${\bm J}$ has the form 
\begin{align}
k(\bm{J}, \bm{J})
&= \notag
\big\<
\bm{J} | k | \bm{J}
\big\>
= 
{\rm cov} \big(
\<\bm{J}, S\>, \<\bm{J}, S\>
\big)
\\ 
&= (\nabla_{\bm p} \mathcal{K} )^{-1} {\rm cov}\big(\<\nabla_S \mathcal{K}, S\>, \<\nabla_S \mathcal{K}, S\>)
\big) (\nabla_{\bm p} \mathcal{K} )^{-T}.
\label{eq:k_J_J}
\end{align}
The exact components of matrix ${\rm cov}\big(\<\nabla_S \mathcal{K}, S\>, \<\nabla_S \mathcal{K}, S\>) \big)$ is given in Appendix \ref{appd:err_system_of_lin_eq}. The covariance matrix $k({\bm J}, {\bm X})$ from \eqref{eq:var_j} is computed similarly.

The predicted uncertainty can be minimized in an optimal manner via active learning. We introduce the information function of point $X$ in the following way: 
\begin{align}\label{eq:inform_func}
    \mathcal{H}(X) = \sum_i \mathcal{H}(\mathcal{Q}_i|X) = - \sum_i \log \frac{\mathbb{V}[\mathcal{Q}_i| \text{Data} \cup X]}{\mathbb{V}[\mathcal{Q}_i| \text{Data}]},
\end{align}
where $\text{Data}$ is the input data and $\mathcal{Q}_i$ are the linear functionals of the free energy the uncertainty in which we want to reduce. In our study these functionals describe the phase transition points (e.g., on the melting curve). For example, we can fix concentrations of the first phase $c_1^{(i)}$ and for each of these define the quantity $\mathcal{Q}_i$ as the temperature of transition, $\mathcal{Q}_i = T\big(c_1^{(i)}\big)$, where $T$ is the solution of \eqref{eq:sys_eq_1}--\eqref{eq:sys_eq_2}, and $\mathbb{V}[\mathcal{Q}_i| \text{Data} \cup X]$ and $\mathbb{V}[\mathcal{Q}_i| \text{Data}]$ is computed with the help of \eqref{eq:var_j}.

Our learning policy is to perform a simulation at the new point $X_*$ given by 
\begin{align*}
    X_* = \argmax_{X} \mathcal{H}(X).
\end{align*}
The greedy strategy is then iteratively repeated until reaching the desired convergence. 

\begin{figure}
	\includegraphics[width=8cm]{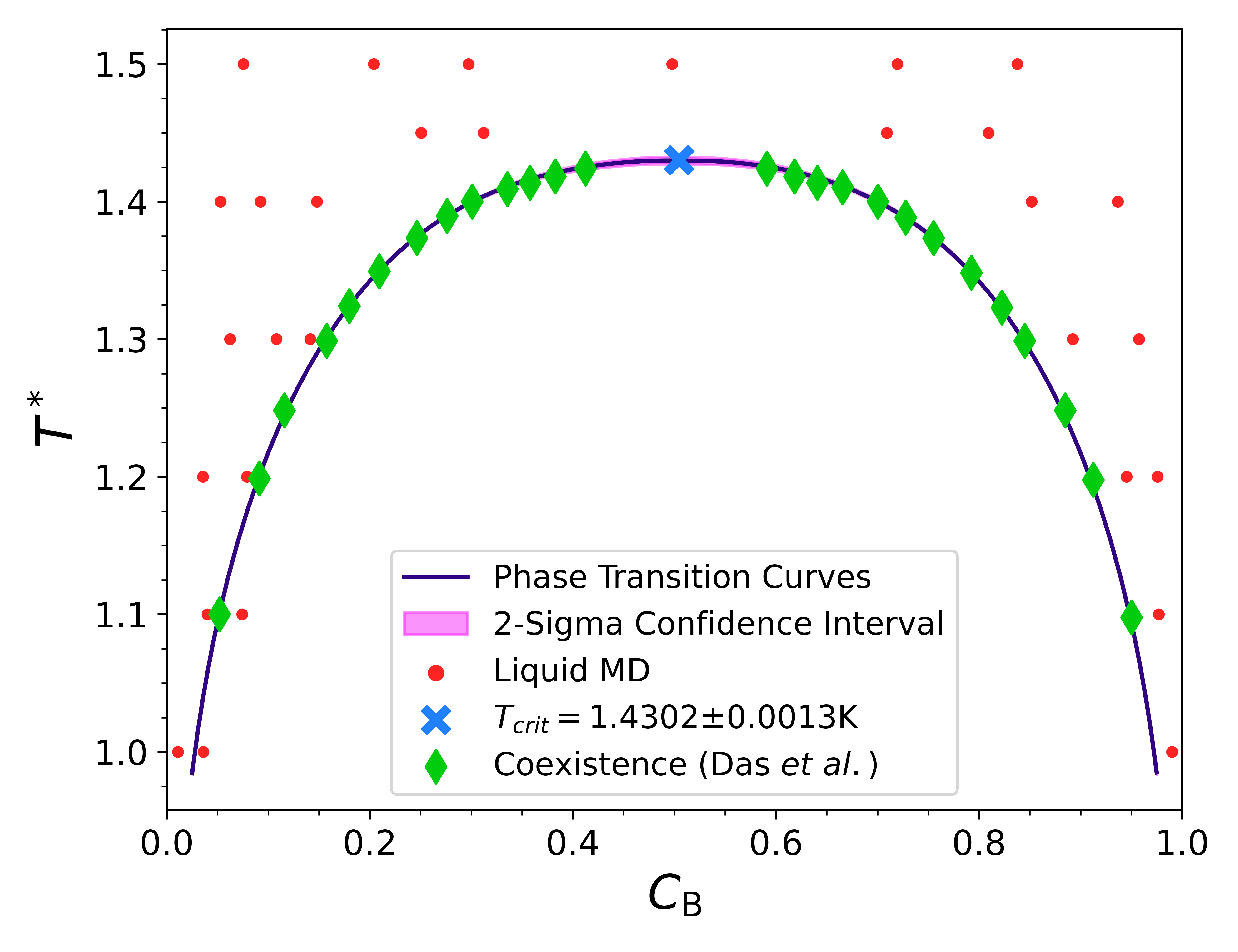}
    \caption{The phase transition curve with the two-sigma confidence interval of the symmetrical truncated Lennard-Jones binary liquid for $N=6400$ atoms. The red points denote the MD simulations we used to reconstruct the liquid free energy. Our algorithm predicts the critical temperature $T_{\rm crit} = 1.4302 \pm 0.0013 $. The predicted transition curves are in good agreement with the coexistence simulations reported by Das \textit{et al.}\ in Ref.~\cite{binary_lennard_jones}.}
    \label{fig:lj_seg}
\end{figure}

\section{Results}
\label{sec:results}

\subsection{Validation on the Lennard-Jones system}
We begin by validating our algorithm on the binary Lennard-Jones system which has previously been studied in \cite{binary_lennard_jones}. We adopt the same simulations conditions as those in \cite{binary_lennard_jones}, namely we utilize the truncated Lennard-Jones potential \cite{Allen_Tildesley_book} to describe the interaction between the type-A and type-B particles. The potential parameters are set as follows: $\sigma_{AA} = \sigma_{BB} = \sigma_{AB} = \sigma$, $\epsilon_{AA} = \epsilon_{BB} = 2 \epsilon_{AB} = \epsilon$ and $r_c = 2.5 \sigma$. We fix the reduced density $\rho^* = \rho \sigma^3 = 1 $ and the total number of atoms $N=6400$. The reduced temperature $T^* = T/\epsilon$ lies within the interval $[1, 1.5]$.

Figure \ref{fig:lj_seg} presents the obtained results. The red points represent the data selected for the simulations. We chose a total of 30 points, and at each point we performed eight simulation consisting of 20\,000 times steps each. By learning the data, our algorithm produced the phase segregation curves with a two-sigma confidence interval.

The coexistence points obtained by Das \textit{et al.}\ in Ref.~\cite{binary_lennard_jones} are in good agreement with our segregation curve. The critical point obtained by our algorithm is $T_{\rm crit} = 1.4302 \pm 0.0013$ which agrees with the value of $T_{\rm crit} = 1.423 \pm 0.002$ as reported in \cite{binary_lennard_jones} which was obtained by fitting the coexistence points within the range $0.2 < C_{B} < 0.5$ (but excluding the two points closest to $T_{\rm crit}$).

We next proceed with applying our algorithm to real systems modeled with a machine-learned potential fitted to ab initio data.

\subsection{Ge--Si}
We start by applying our algorithm to the Ge--Si system.
Its phase diagram is relatively simple, has no stable intermetallic phases and has the diamond-Ge phase and diamond-Si phase featuring a miscibility gap below a certain critical temperature and a solid solution above this temperature, which melts with further raising the temperature.
We thus will study how well the miscibility gap and the solidus/liquidus lines are captured with our method.

We used the moment tensor potentials (MTP) \cite{shapeev2016-mtp} with active learning to construct the interatomic potential in an automatic manner.
To that end, we run molecular dynamics simulations with different values of chemical potential and different temperatures covering the entire phase diagram while actively training the potential on-the-fly \cite{Podryabinkin2017,gubaev2019-alloys}.
A total of 578 configuration were selected while constructing the MTP potential, all of which were computed with density functional theory (DFT).
We used the VASP software for the DFT calculations with the GGA-PBE density functional and PAW pseudopotentials with four valence electrons \cite{VASP1,VASP3,VASP4}.
The \texttt{ENCUT} parameter was chosen as 450 eV which is $1.8\cdot\max(\texttt{ENMAX\_GE}, \texttt{ENMAX\_SI})$, where \texttt{ENMAX} is the energy cutoff of the corresponding pseudopotential, and 3x3x3 k-point mesh centered at the gamma-point was used for 64 atoms to ensure convergence of DFT computations up to 1 meV/at.  

Before starting our Bayesian regression algorithm, we computed the melting points of Ge and Si and obtained the following results: $T_{\rm m}({\rm Ge}) =  931.05 \pm 0.91 K$ and $T_{\rm m}({\rm Si}) = 1460.62 \pm 0.95 K$.

% We next conduct the phonon calculations with MTP to obtain the $\log\det H$ part of the entropy for the Si and Ge structures at $T=0$.
% We do it with $N=...$...

After these preparations, we applied our Bayesian regression algorithm to construct the Ge--Si phase diagram. We conducted simulations for $T$, $\mu$, $N$ as follows. We first learned the dependence of $c = c(T, \mu, N)$ using the supplementary dataset with points ($T$, $\mu$, $N$) randomly chosen in the following ranges: $T \in [0 K, 2000 K]$, $\mu \in [-20, 20]$, $N \in \{216, 512, 1000, 1728\}$. We used short MD trajectories for the supplementary dataset as we aim to approximately estimate $c = c(\mu, T, N)$. Then, we selected 50 points for the crystalline phase and 28 points for the liquid phase that nearly uniformly covered the $T$-$c$-$N$ region.
The selected points were computed with 10 times longer MD trajectories (the total of 6 independent trajectories with 20 000 time steps of 1fs each) and fed as input data to our algorithm. We discarded points that underwent phase transition during the simulation, which we identified by the polyhedral template matching algorithm \cite{Larsen_2016}. 

\begin{figure*}
	\includegraphics[width=15cm]{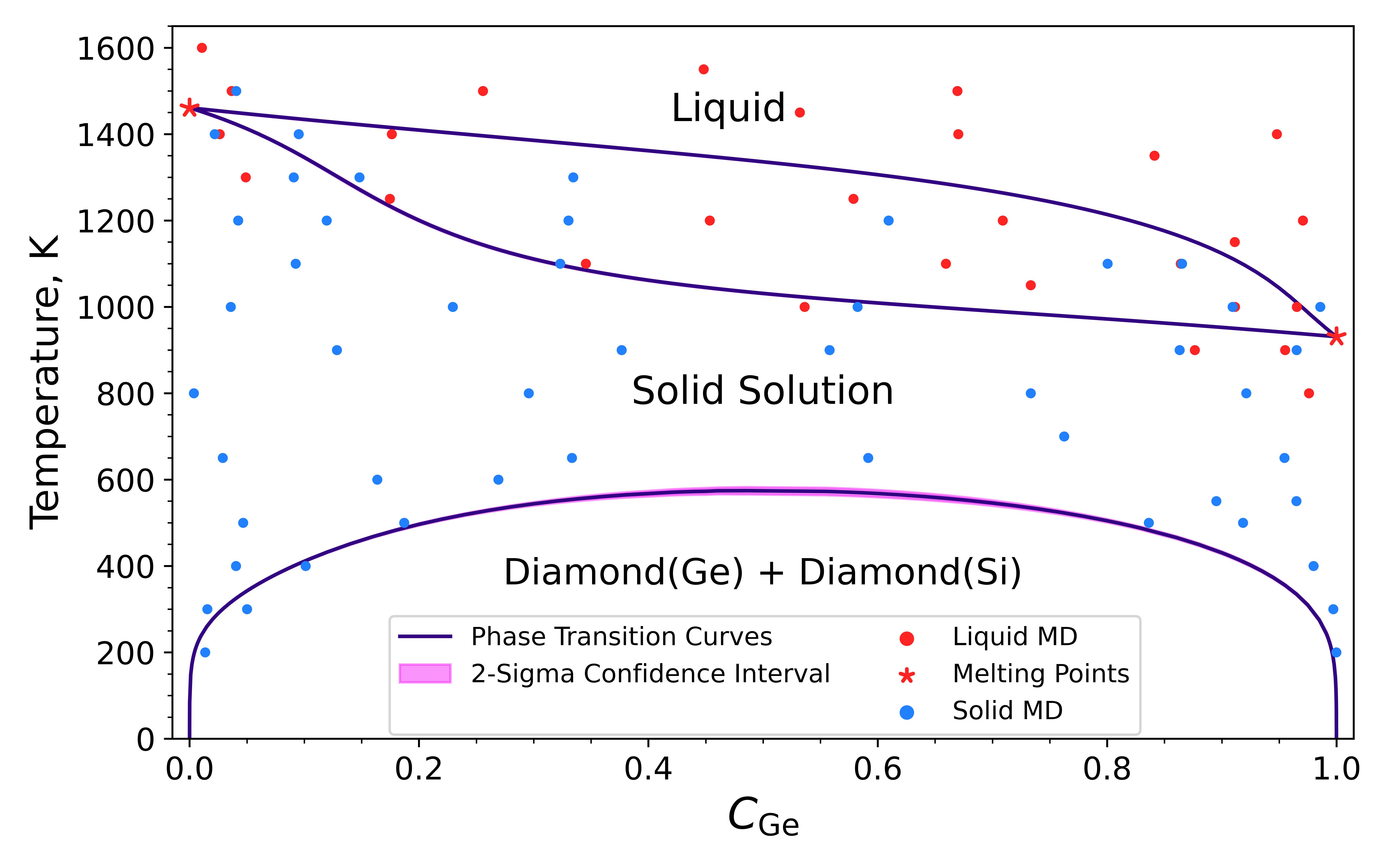}
    \caption{Ge--Si phase diagram. The phase transition lines are plotted with the two-sigma confidence interval which is of the order of 2$ K$. Red markers denote the data (MD simulations, melting points) we used to reconstruct the liquid free energy, while the blue markers refer to the data used for the solid phase.
    	Phonon simulations are not necessary as we only need a relative additive shift between the free energies of solid solution and liquid.
    	We found the critical temperature to be $T_{\rm crit} = 575 \pm 4 K$ at the concentration $c_{\rm crit} = 0.498 \pm 0.006$.}
    \label{fig:gesi}
\end{figure*}
 
%\begin{figure}[h!]
%	\centering
%	\includegraphics[width=8cm]{ge_si_seg.jpg}
%    \caption{Segregation of Ge--Si. Walle et al \cite{Walle_2002} do not consider the vibrational contribution to the entropy.}
%    \label{fig:gesi_seg}
%\end{figure}

\begin{figure*}
	\includegraphics[width=15cm]{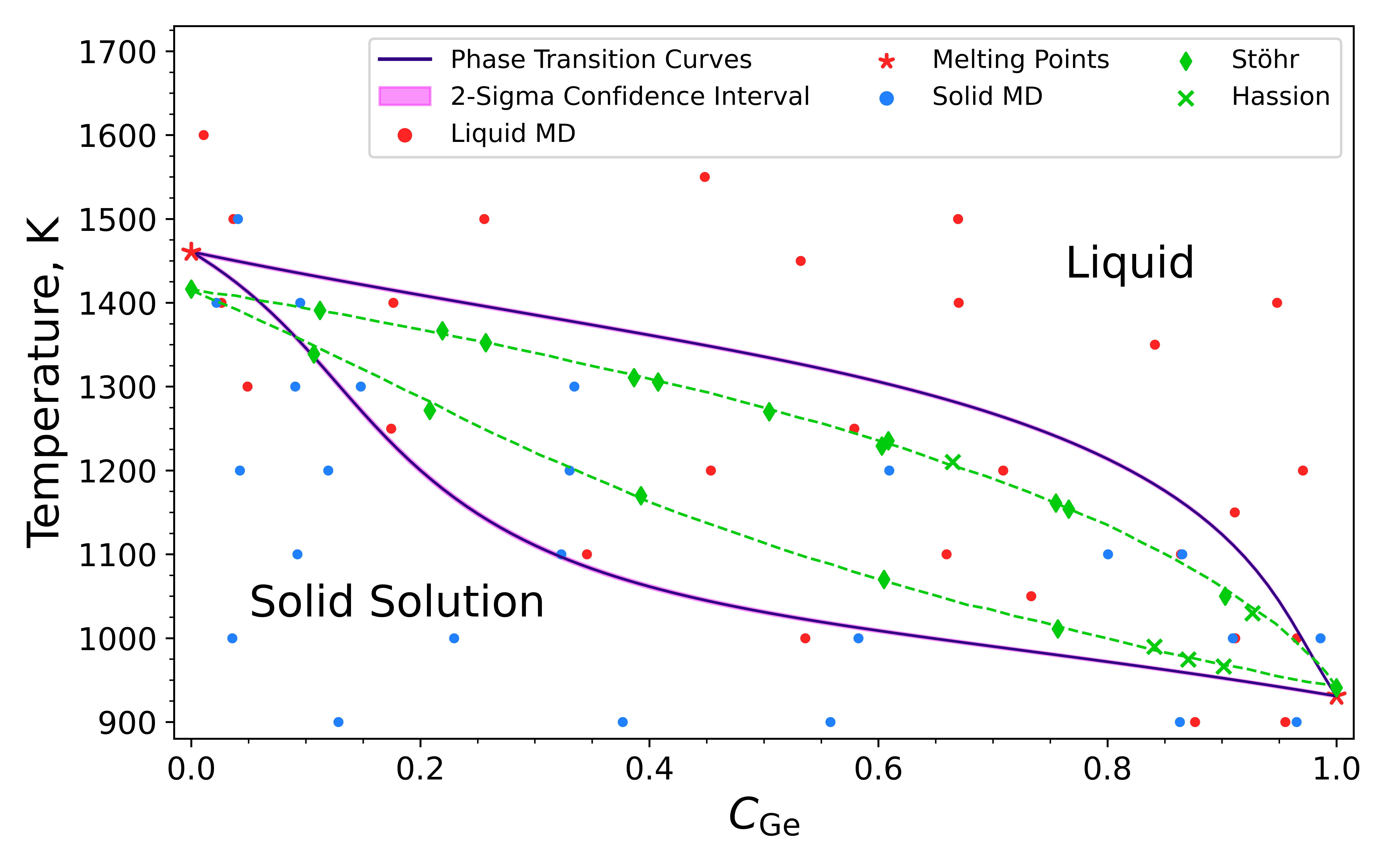}
    \caption{Melting curves of the Ge--Si system. We shift the experimental data points  --- denoted as Stöhr \cite{Stohr_1939} and Hassion \cite{Hassion_1955} --- down by 270 $ K$ for a visually better comparison (it is known that DFT with the PBE functional typically underpredicts the melting points by 100--300$ K$).}
    \label{fig:gesi_melting}
\end{figure*}

We conducted a numerical experiment to validate our methodology:
we ``hid'' the melting point of Si, keeping only the melting point of Ge in the training dataset, then reconstructed free energies of both phases and thus predicted the melting point of Si which was then compared to the value from the coexistence simulations.
We note that a single data point (namely, the Ge melting point) is sufficient to determine the additive shift of the liquid free energy relative to the solid free energy, however, this test constitutes a significant challenge, because the Gaussian process needs to accurately ``integrate'', in its data-driven unstructured-mesh manner, the datum at $c_{\rm Ge}=1$ all the way to $c_{\rm Ge}=0$ using only derivatives of the free energy. We found that the melting temperature predicted by our algorithm  $T_{\rm m} = 1461.31 \pm 1.82 K$ lies within the one-sigma confidence interval of the melting point obtained by the coexistence simulations ($T_{\rm m}({\rm Si}) = 1460.62 \pm 0.95 K$). This validates our method and also demonstrates its high efficiency.

% The algorithm produced the phase diagram of Germanium--Silicon over its entire existence range, along with the two-sigma confidence interval for the phase transition curves for infinite atoms in the system ($N \rightarrow \infty$). The diagram is plotted in Figure Figure \ref{fig:gesi}, which shows that our algorithm qualitatively reproduced all its features. 

Figure \ref{fig:gesi} shows the phase diagram produced by our algorithm for the Ge--Si system over its entire existence range, along with the two-sigma confidence interval for infinite atoms in the system ($N \rightarrow \infty$). The algorithm qualitatively reproduced all the features of the diagram.
As can be seen from Figure \ref{fig:gesi_melting}, the melting points are underpredicted by about 300$ K$ as compared to the experimental values, which is typical for the PBE exchange-correlation functional.
The solidus-liquidus gap is slightly wider than that from the experiment \cite{Olesinski1984}, nevertheless is well-reproduced qualitatively..
As for the miscibility gap, we found no experimental data to compare with.
In \cite{Walle_2002}, Monte-Carlo simulations of the miscibility gap was performed with a cluster-expansion model and a value close to 300$ K$ was reported, whereas in our study a gap of about 600$ K$ was obtained.

The total computational effort spent on constructing 
the Ge--Si phase diagram was approximately $23\,000$ CPU-hours, out of which the DFT calculation of the training set required $2\,000$ CPU-hours, the melting point calculations also took about $2\,000$ CPU-hours, and the rest was spent on running the $\mu$pT-MD simulations.

\subsection{K--Na}

\begin{figure*}
	\includegraphics[width=15cm]{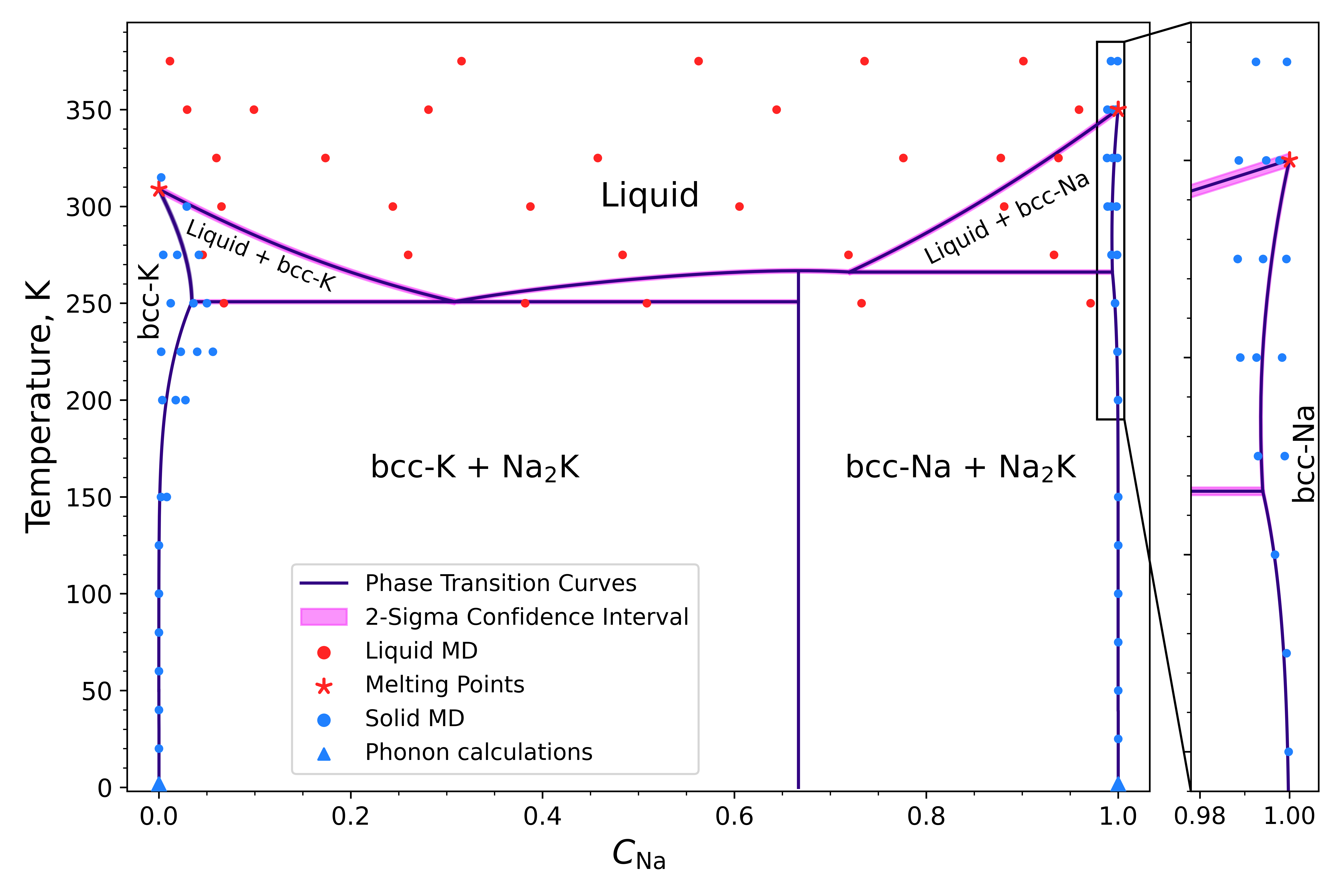}
    \caption{K--Na phase diagram with the two-sigma confidence interval. Red markers denote the data (MD simulations, melting points) we used to reconstruct the liquid free energy, while the blue markers (MD simulations, phonon calculations) refer to the data used to learn the free energy of the bcc phase.}
    \label{fig:kna_full_gp}
\end{figure*}

We next test our algorithm on the K--Na phase diagram.
We chose the K--Na system for two reasons. The first reason is that this phase diagram, while still relatively simple, contains a number of nontrivial features that constitute a challenging test for our algorithm. There are two separate body-centered cubic (bcc) phases, namely bcc-Na and bcc-K, which are not connected in a single solid solution phase, an intermetallic phase, and several regions where the solid and liquid phases coexist. Additionally, an eutectic point is present, at which both Na$_2$K and bcc-Na solid phases simultaneously solidify.
The second reason is purely methodological: both K and Na do not have valence d-electrons and therefore are rather accurately modeled with DFT---we will see that without any temperature shift the phase diagram is quantitatively close to the experimental one.

To that end, we chose accurate pseudopotentials for K and Na, with nine valence electrons each.
As in the previous example, we used the VASP software for the DFT calculations with the GGA-PBE density functional and PAW pseudopotentials \cite{VASP1,VASP3,VASP4}.
We performed several tests to ensure the convergence of the DFT calculations with respect to the k-points and cutoff energy to construct a robust interatomic potential. For k-points, we found that the convergence up to 1 meV/at for 54 atoms is reached on the 4x4x4 grid centered at the gamma-point. We determined that in order to achieve the virial stress convergence up to 1 meV/at, it is necessary to employ a plain wave energy cutoff of 750 eV, which equals to $2\cdot\max(\texttt{ENMAX\_NA}, \texttt{ENMAX\_K})$.

The moment tensor potential for the K--Na system was constructed similarly to the Ge--Si system.
The training set consisted of 325 configurations sampled randomly from molecular dynamics and additional 161 configuration acquired during active learning. 

We next derive the asymptotic free energy for the ${\rm Na}_2{\rm K}$ intermetallic phase (Appendix \ref{app:asym_im}). The elementary cell of ${\rm Na}_2{\rm K}$ consists of 12 atoms, yet there are only three distinct substitutional defects possible: K substitutes Na with the formation energy $E_{1}$ (6 sites) and $E_{2}$ (2 sites), while Na substitutes K with the energy  $E_3$ (4 sites).
The asymptotic free energy for ${\rm Na}_2{\rm K}$ has the form
\begin{align*}
	&\vspace{-1em}\textstyle
	G^{\rm im} (T, c_{\rm Na}, N) =
	E_0 - {\textstyle \frac32} T \log(2\pi T) + {\textstyle \frac12} T N^{-1} \log\det \hat{H}_0
	\\ &~\textstyle
	- \frac{\sqrt{2}}{3} \sqrt{e^{\beta  E_1}+3 e^{\beta  E_2}} e^{-\frac{1}{2} \beta  (E_1 + E_2 + E_3)}
	\\ &~\textstyle
	+ 
	\frac{1}{2}  \left(\log \left(\frac{e^{\beta  E_1}+3 e^{\beta  E_2} }{2} \right)-\beta  (E_1 + E_2 - E_3)\right) \left(c_{\rm Na} - \frac{2}{3}\right)
	\\&~\textstyle
	+ 
	\frac{3}{2 \sqrt{2}} \frac{e^{\frac{1}{2} \beta  (E_1 + E_2 + E_3)}}{\sqrt{e^{\beta E_1}+3 e^{\beta E_2}}} \left(c_{\rm Na} - \frac{2}{3}\right)^2.
\end{align*} 

We applied the same methodology as for the Ge--Si system to collect the dataset for each phase: we randomly sampled $(T, \mu, N)$ to learn how $c$ depends on $(T, \mu, N)$ and then generated the $(T, \mu, N)$ points trying to uniformly cover the T-c-N region, where $N \in \{128, 250, 432, 686, 1024, 1458\}$.
In total, we acquire 45 points for the bcc phase and 31 points for the liquid phase.
For the ${\rm Na}_2{\rm K}$ intermetallic phase we did not sample any points; instead, we simply took the asymptotic free energy of ${\rm Na}_2{\rm K}$ as its free energy.  

We next conducted phonon calculations with the MTP potential to obtain the $\log\det \hat{H}_0$ part of the entropy for the K, Na, and ${\rm Na}_2{\rm K}$ structures at $T = 0 K$. We implemented the calculation of the energy Hessian from scratch in Python \cite{python_citation} using a second-order finite-difference scheme; the evaluation of the forces for intermediate configurations was performed using the MLIP-2 software package \cite{novikov2020-mlip}.
In our algorithm, $\log\det \hat{H}_0$ serve to determine the additive shifts of the solid free energies.
We performed calculations with $N \in \{128, 250, 432, 686, 1024, 1458\}$ for bcc phases and with $N \in \{ 48, 384, 1296, 3072, 6000\}$ for ${\rm Na}_2{\rm K}$ and then fitted $\log\det \hat{H}_0 (N)$ using Gaussian process regression to obtain the values as $N \rightarrow \infty$. The melting points of Na and K provide the additive shifts between the bcc and liquid phases that we used to reconstruct the free energy of the liquid phase. The melting points for the unary structures were $T_{\rm m}({\rm K}) = 309.52 \pm 0.92 K$, $T_{\rm m}({\rm Na}) = 349.60 \pm 0.88 K$. 

We performed a similar validation test as we did for Ge--Si: we ``hid'' the melting point of Na and then reconstructed the phase diagram and melting point of Na.
Without the melting point of Na as an input datum, the algorithm functioning can be described as follows.
It effectively ``integrates'' the free energy of bcc-Na from $(T,c)=(0,1)$ to $(T_{\rm m}({\rm Na}),1)$ and independently integrates the free energy of bcc-K and liquid from $(0,0)$ to $(T_{\rm m}({\rm K}),0)$ and then to $(T_{\rm m}({\rm Na}),1)$ at which point it compares the free energies of bcc-Na and liquid.
(This is, of course, a simplified description---in reality the algorithm finds the unknown $T_{\rm m}({\rm Na})$ and performs the integration at the same time.)
Thus predicted melting temperature $T_{\rm m}({\rm Na}) = 354.1 \pm 2.4 K$ coincided with the value obtained by coexistence simulations ($T_{\rm m}({\rm Na}) = 349.60 \pm 0.88 K$) within two-sigma.
%This validates accounting of phonon calculation in our algorithm.

\begin{figure}[h!]
	\centering
	\includegraphics[width=8cm]{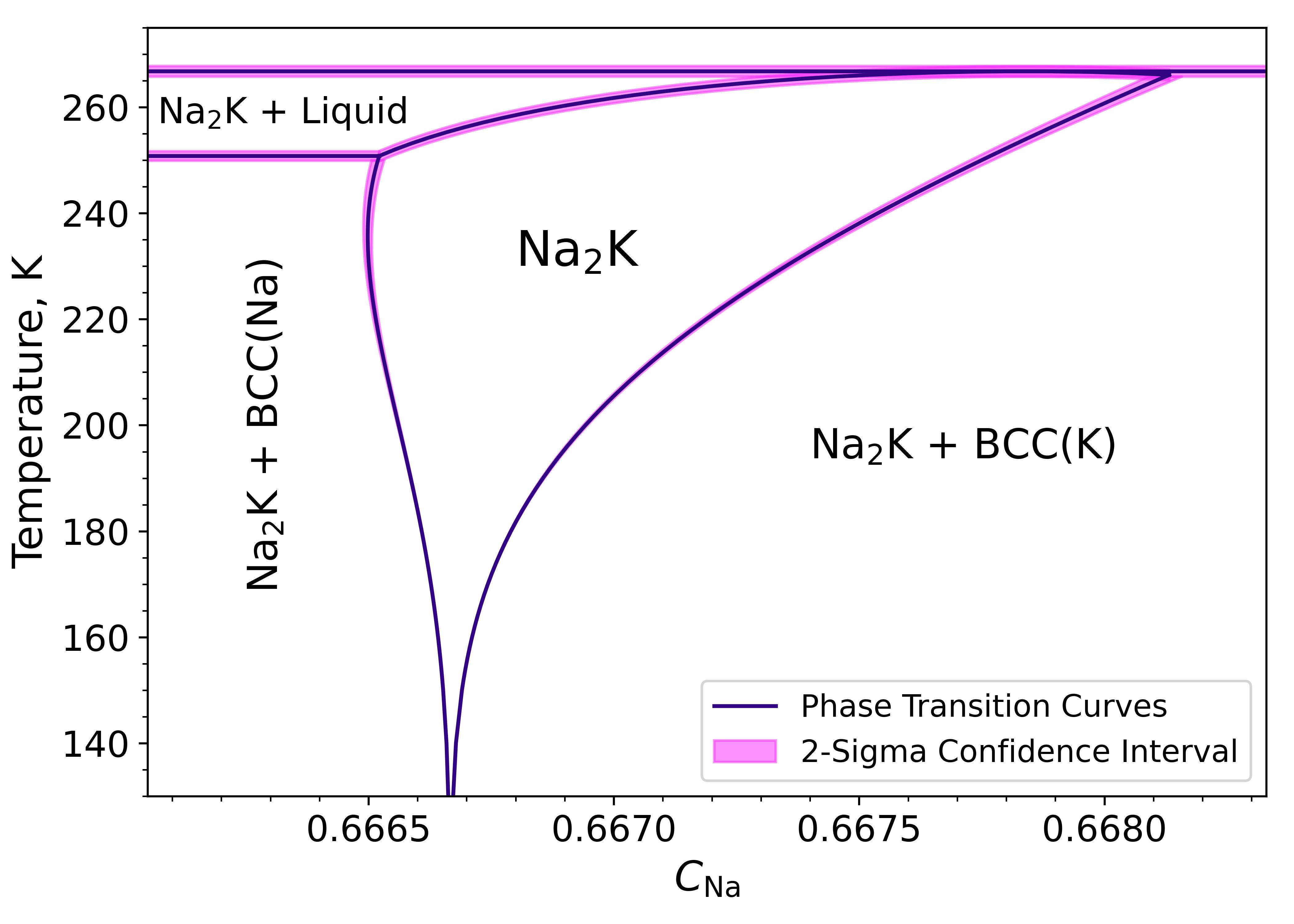}
    \caption{Our algorithm, in principle, allows for a detailed resolution of the widening of intermetallic phases; the figure illustrates the phase diagram details for concentrations near the Na$_2$K phase.
    }
    \label{fig:kna_im_loop_gp}
\end{figure}

\begin{figure*}
	\includegraphics[width=15cm]{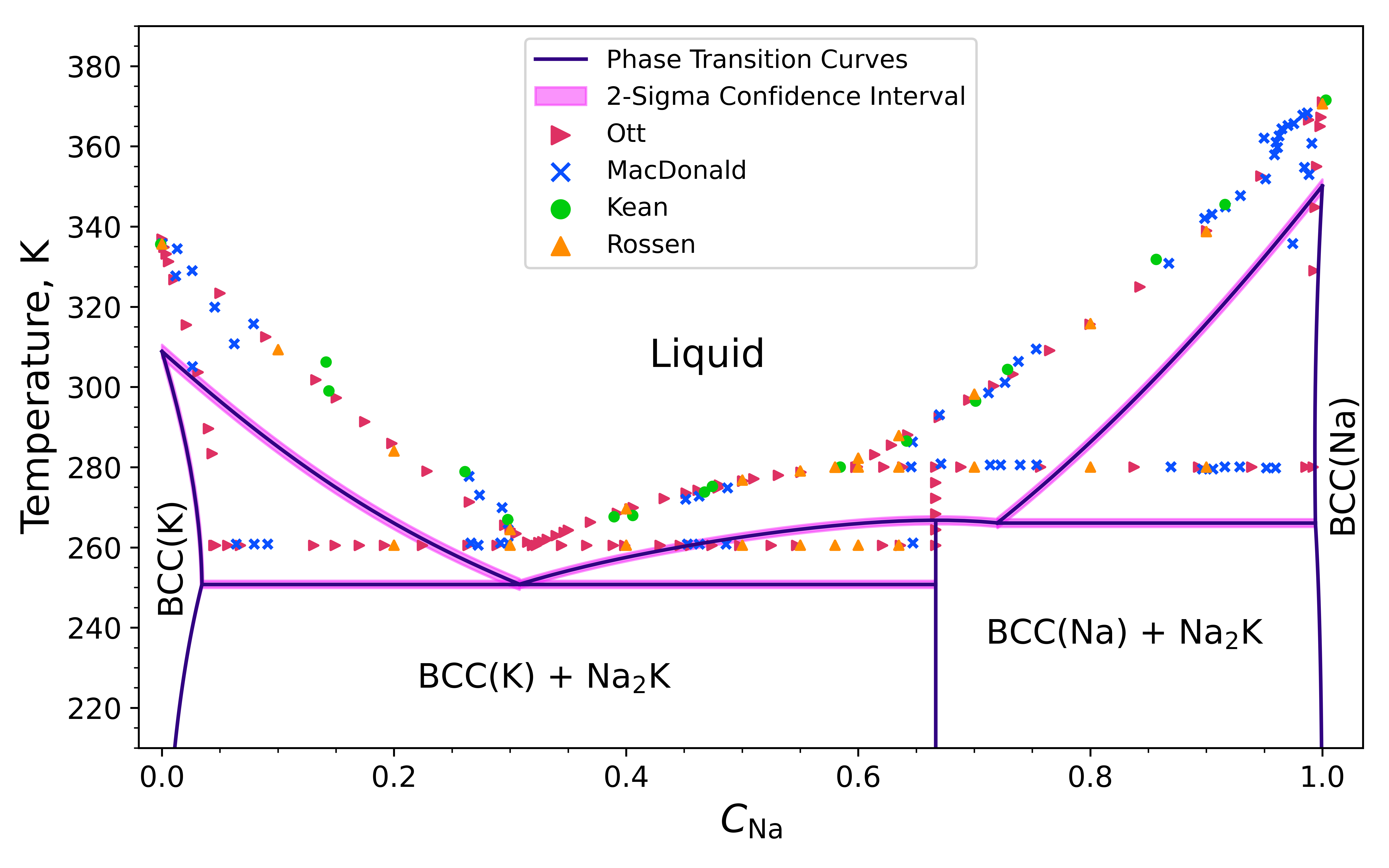}
    \caption{Computed K--Na phase diagram is compared to the experimental data: Ott \cite{ott_69}, MacDonald \cite{MacDonald_56}, Kean \cite{Kean_39}, Rossen \cite{Rossen_1912}. 
    Our algorithm quantitatively reproduces all the features of the phase diagram (within about 20$ K$ in temperature and $10^{-2}$ in concentration), with the exception of the triple point of ${\rm Na}_2{\rm K}$--Na--liquid coexistence which is shifted by about $0.12$ in concentration away from the Na-rich side.}
    \label{fig:kna_exper1}
\end{figure*}

Figure \ref{fig:kna_full_gp} presents the phase diagram obtained with our algorithm with the two-sigma confidence interval for infinite ($N\to+\infty$) atoms in the system for the entire range of temperatures and concentrations. 
The red and blue markers denote the input data for liquid and bcc phases that we used in our algorithm.  
%We observed that the $\mu$pT-simulations of the bcc phase become unstable when performed away from Na-rich and K-rich regions, although the available data was sufficient for our algorithm to learn the free energy of bcc phase and reconstruct the phase diagram.  
In Figure \ref{fig:kna_im_loop_gp}, we provide the detailed plot around the critical point of the intermetallic phase.
Although the sub-percent resolution in concentration of phase diagrams might not be required in practice, we still emphasize that our algorithm is capable of resolving such details.

Figure \ref{fig:kna_exper1} compares the results with the available experimental data \cite{ott_69,MacDonald_56,Kean_39,Rossen_1912}.
Most of the features of our phase diagram quantitatively agree with the experimental studies (within about 20$ K$ in temperature and $10^{-2}$ in concentration), with the exception of the triple point of ${\rm Na}_2{\rm K}$--Na--liquid coexistence which is shifted by about $0.12$ in concentration away from the Na-rich side.

The computational effort of constructing the K--Na phase diagram was approximately  equal to $131\,500$ CPU-hours, most of which, $125\,000$ CPU-hours was spent on the DFT calculation of the 486 configurations (the training set of the MTP potential), which was computationally demanding due to the high energy cutoff of the plane-wave basis.
The melting point calculations took about $300$ CPU-hours and the MD simulations took about $6\,000$ CPU-hours.

\subsection{Active learning}

The confidence intervals of the phase diagram can be minimized by adding new points to the dataset in an optimal way.
To that end, we utilize the information function introduced in \eqref{eq:inform_func}.
The numerator inside the logarithm incorporates the expected variance of the quantity $Q_i$ after adding a new point $X$ into the dataset. 
When computing this variance, we assumed that $\<E\>$ and $\<c\>$ would be computed with a statistical error about one order of magnitude lower than the typical error.
The policy is then to add $X_*$ which maximizes the information function.

\begin{figure*}
    \begin{tabular}{c c c c}
    & {\large step 1} & {\large step 2} & \\
     \parbox[t]{3mm}{\rotatebox{90}{\hspace{22mm} {\large liquid}}} & 
     \includegraphics[width=0.3\textwidth]{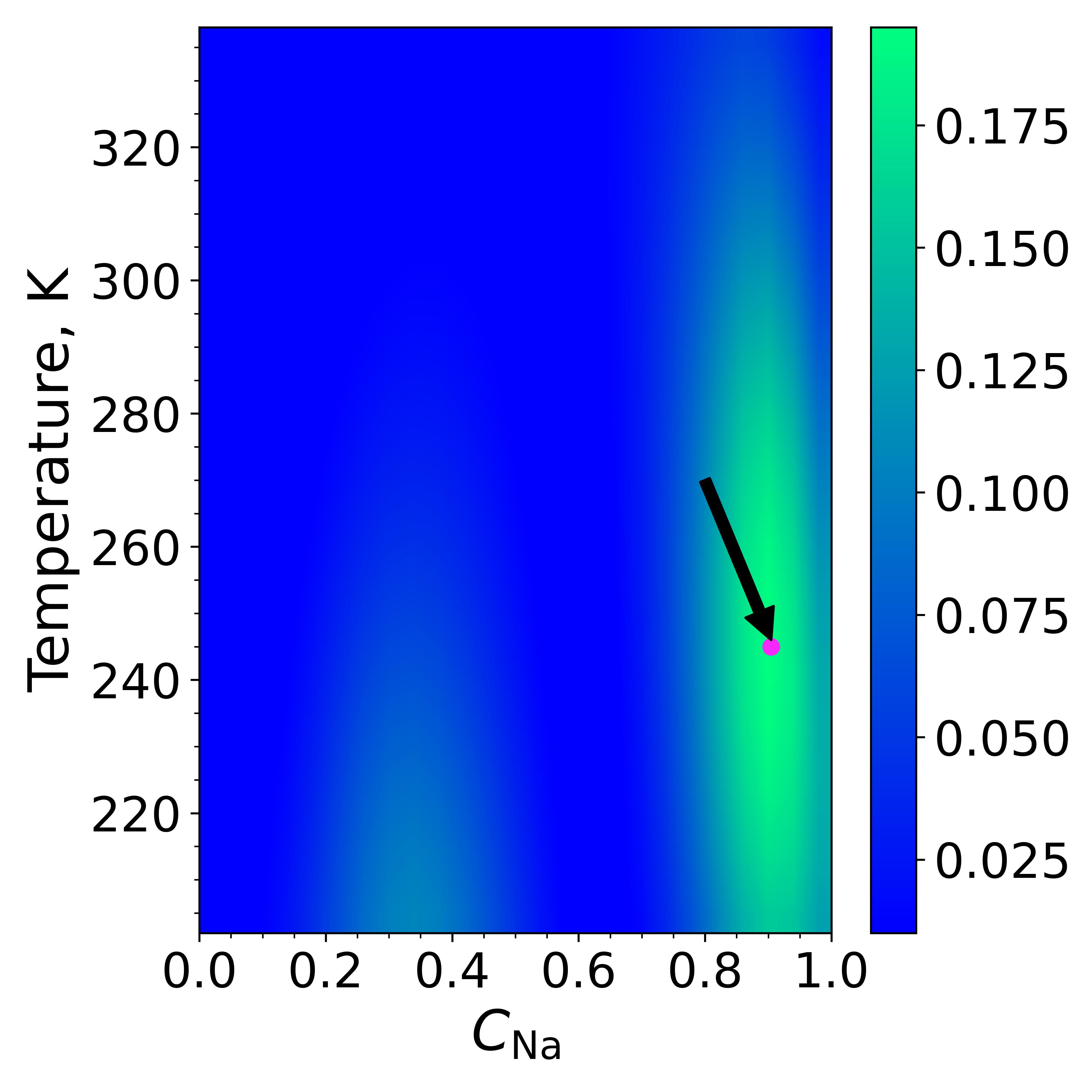} & \includegraphics[width=0.3\textwidth]{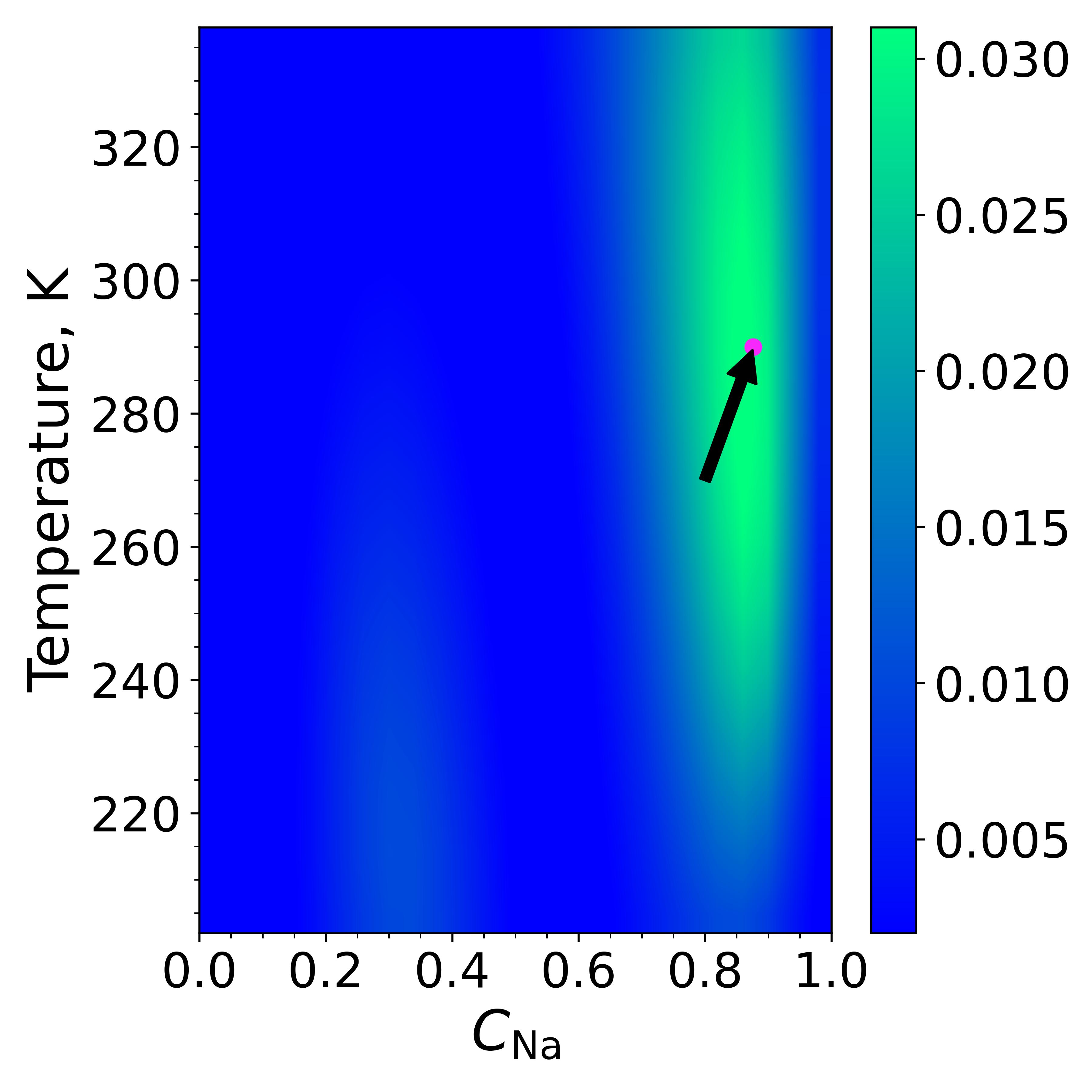} & \includegraphics[width=0.3\textwidth]{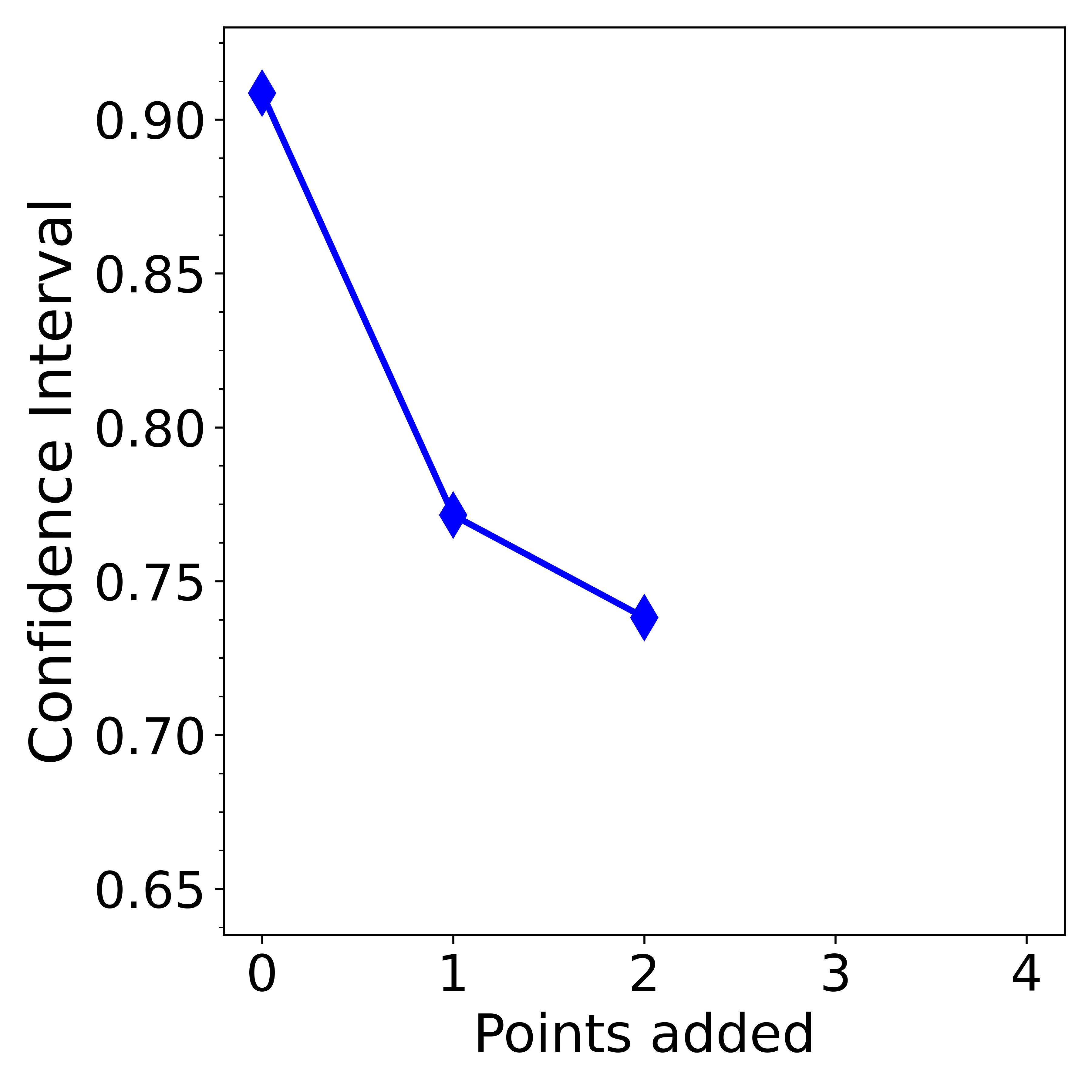}\\
     & {\large step 3} & {\large step 4} & \\
     \parbox[t]{3mm}{\rotatebox{90}{\hspace{25.3mm} {\large bcc}}} & \includegraphics[width=0.3\textwidth]{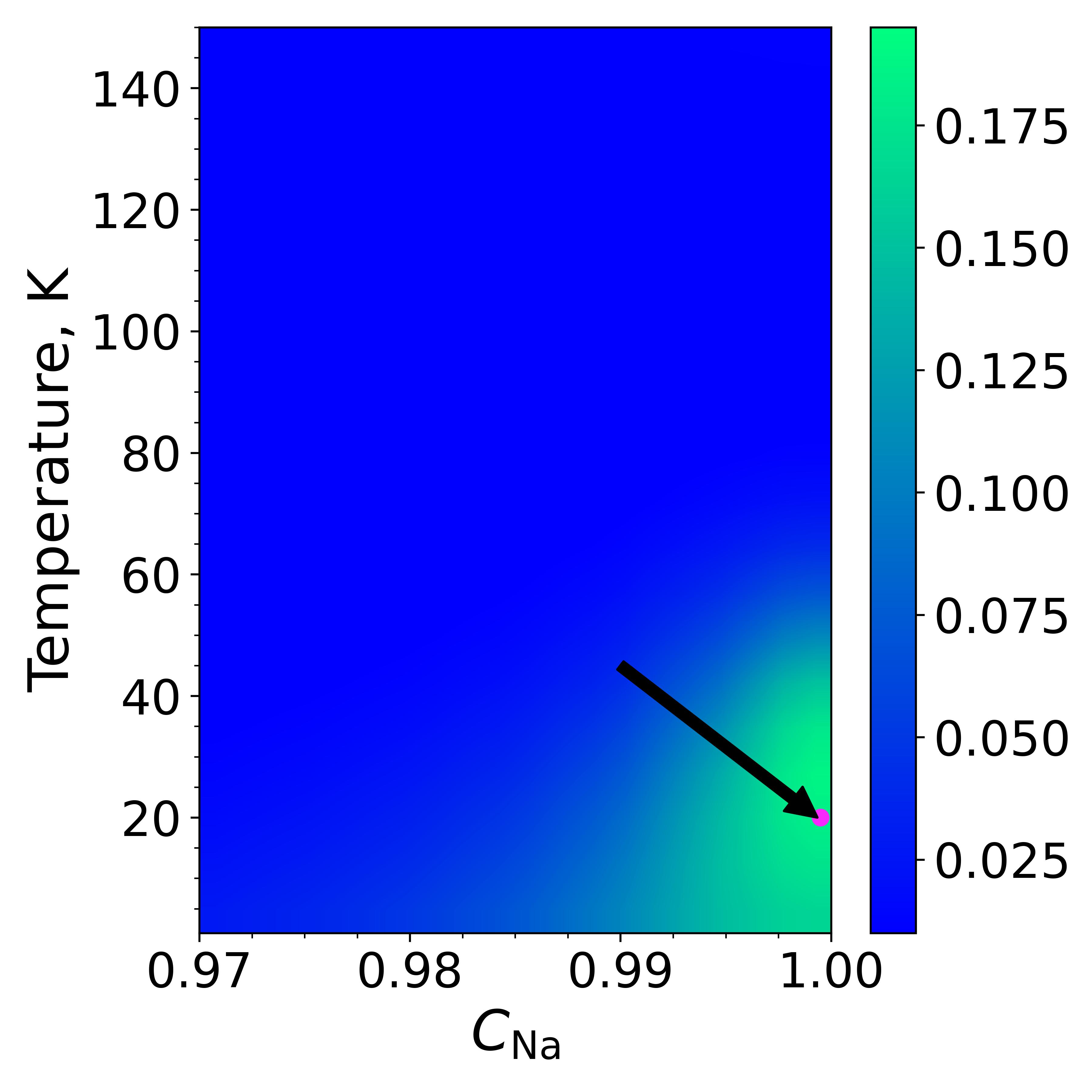} & \includegraphics[width=0.3\textwidth]{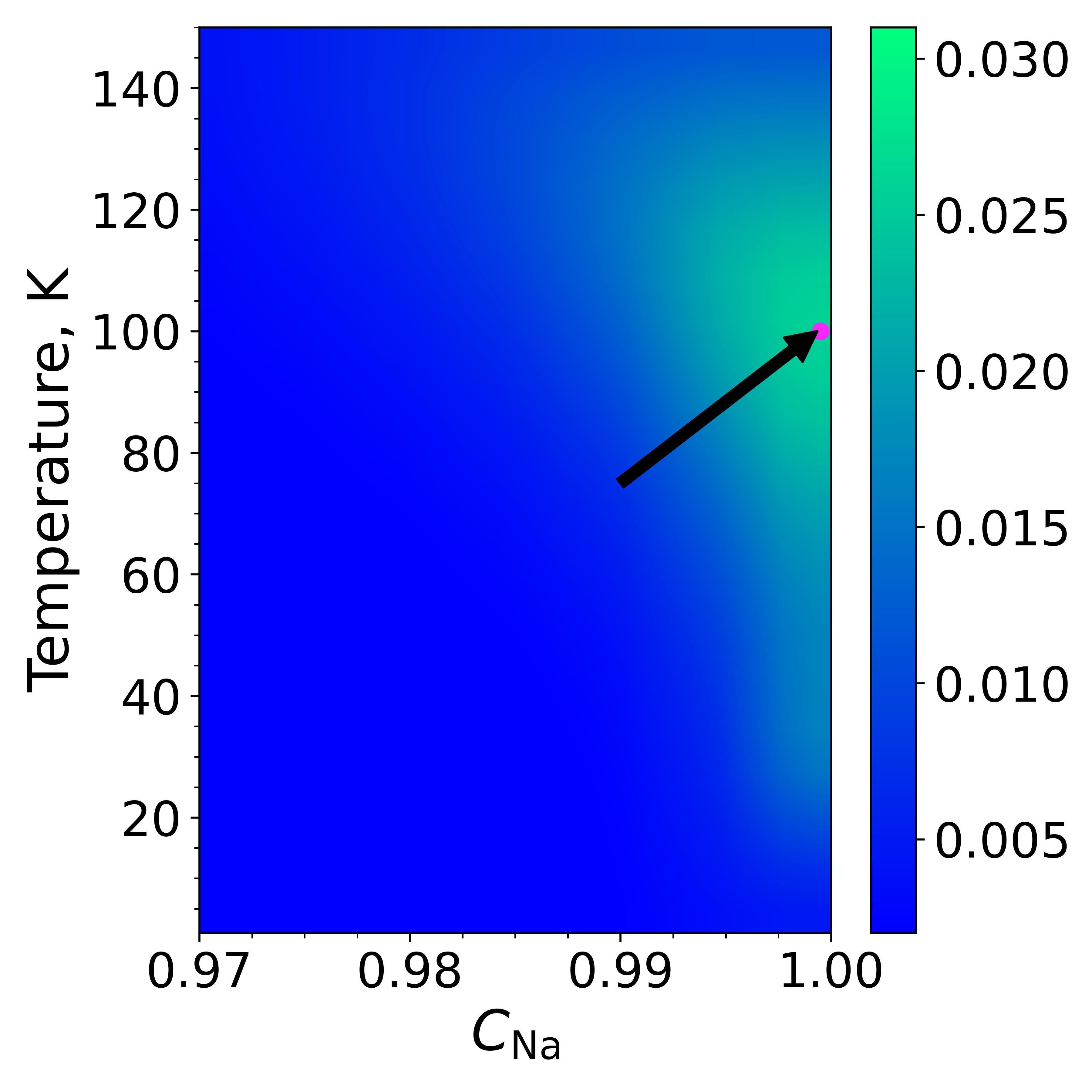} & \includegraphics[width=0.3\textwidth]{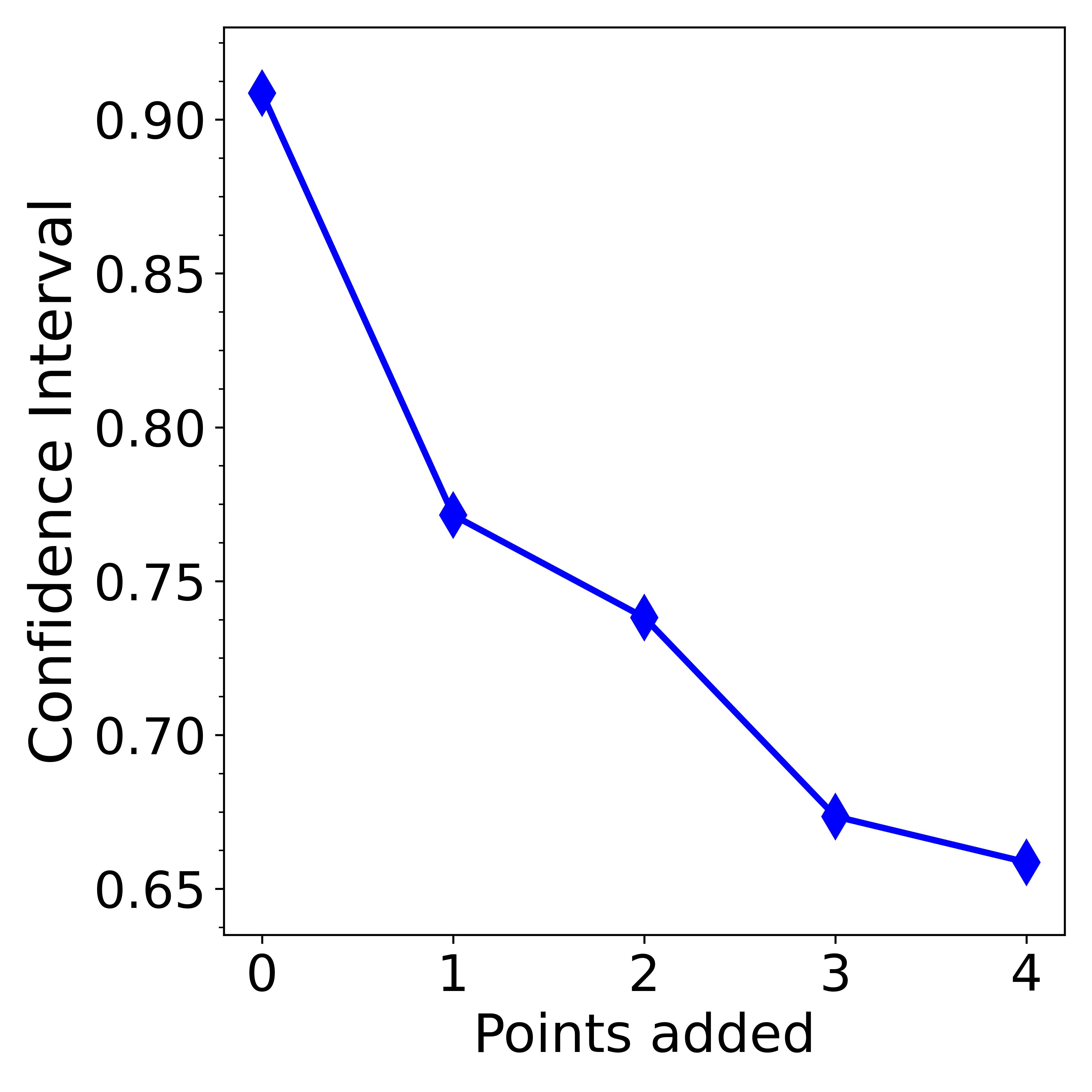}\\
    \end{tabular}
    \caption{The confidence interval of the triple point ${\rm Na}_2{\rm K}$--Na--liquid coexistence was refined from $0.91  K$ to $0.66 K$ via active learning. The density plots (first and second columns) show the information function obtained at each step of active learning for the liquid (first row) and bcc (second row) phases, while the graphs (third column) illustrate how the confidence interval improved after new points $X_*$ were added to the dataset.}
    \label{fig:active_learning}
\end{figure*}

We applied active learning to minimize the confidence interval of the triple point of the ${\rm Na}_2{\rm K}$--Na--liquid coexistence; the data acquisition process is shown in Figure \ref{fig:active_learning}. The first and second columns show the information function. The first row corresponds to the liquid phase while the second row illustrates the information function for bcc-Na. In step 1, the point $X_* = (245 K, 0.905)$ of the liquid phase gave the maximum amount of information, therefore we performed the MD simulation at this point and add it to the dataset. The top right plot shows how the confidence interval is minimized after adding the point $X_*$. The greedy strategy is iteratively repeated until reaching the desired convergence. For example, four points acquired by the active learning policy minimized the confidence interval of the triple point from $0.91 K$ to $0.66 K$. The stopping criterion for the active learning may be chosen in different manner \cite{pmlr-v108-ishibashi20a, Ishibashi2021StoppingCF, Bayessian_optimization}; we leave the selection of the specific criterion for the further investigation.

\section{Summary and Concluding Remarks}
\label{sec:conclusion}

In this work, we have developed, implemented, and tested the Bayesian inference-based algorithm for the construction of binary phase diagrams from data obtained from atomistic simulations based on machine-learning interatomic potentials.
In this algorithm, we first perform an active training \cite{Podryabinkin2017,gubaev2019-alloys,novikov2020-mlip} of an MTP potential during molecular dynamics simulations at the range of conditions relevant to the phase diagram of interest.
Next, with the trained potential, we perform MD simulations in the semi-grand-canonical ensemble to obtain the average concentration and potential energy, as well as conduct phonon (energy Hessian) and melting point calculations; the MD simulations yield data for the derivatives of the free energy for different phases, while the other calculations are used to find the relative shifts of the free energy surfaces with respect to each other.
The simulation data is then fed to a Gaussian process that reconstructs the free energies of the phases together with their uncertainty and also proposes points on the phase diagram, simulation at which would reduce the predictive uncertainty most efficiently.
Moreover, our method, based on the data for different number of atoms in the simulated system, predicts the phase diagram at the ``infinite'' number of atoms, effectively extrapolating the data to $N_{\rm atoms}\to\infty$ and, in combination with the previous feature, proposes also the number of atoms for the simulation in order to reduce the uncertainty most efficiently.

The developed algorithm was implemented and tested on two binary systems, Ge--Si and K--Na, in the full range of concentrations and temperatures (including the liquid phase).
The main reason for choosing Ge--Si was that it has a stable solid solution phase in the entire concentration range which allowed us to conduct a challenging validation test, which our algorithm successfully passed, in which we fed the Ge melting point to the algorithm and let it ``integrate'' (in its data-driven unstructured-mesh manner) the MD data all the way to the melting point of Si which we validated against the coexistence simulations.
The reason for choosing K--Na was that both K and Na are well-modelled with DFT, in the sense that the both melting points are within tens of Kelvin away from the accurately measured value, has an intermetallic phase, and several phase coexistence regions on the phase diagrams.
The algorithm showed high quantitative accuracy, within 20$ K$ in temperature and about $10^{-2}$ in concentration in determining most features of the phase diagram (to be precise, all but the Na-Na$_2$K-liquid triple point).

The proposed methodology can be combined with the developments in \cite{ladygin2021-phad} for the construction of the concentration-pressure-temperature diagrams.
The method can also be extended over to multielement (more than two) phase diagrams, for which a number of extra methodological developments would be required, e.g., the treatment of binary intermetallic phases in which the third (fourth, etc) component is diluted.
Our method computes accurately (to the extent allowed by the interatomic potential) the vibrational and configurational entropy of phases. 
In the future, we plan to extend our method to electronic and vacancy types of entropy.
And last but not least, we plan to allow for multifidelity data, which would allow for conducting a large number of simulations with an interatomic potential and a small number of simulations with DFT to produce a truly DFT-accurate phase diagram, representing a data-driven analog of the free energy perturbation approach, see e.g., \cite{jung2023-therodynamic-int}.

\begin{acknowledgments}
This work was supported by Russian Science Foundation (Grant No. 23-13-00332, https://rscf.ru/project/23-13-00332/). 
\end{acknowledgments}

\appendix
\section{Crystalline free energy asymptotic}
\label{app:free_en_asym}
Here we derive the asymptotic free energy for the crystalline phases (bcc/fcc/diamond) as $\beta^{-1} = T \rightarrow 0$ for a binary system.
We start with \eqref{eq:Phi} and \eqref{eq:partition-function}:
\begin{align}\label{eq:ap1:phi}
\begin{split}
\Phi &= -T \log \Bigg( \int_{0}^{\infty} \exp(-\beta p \hat{V}) {\rm d}\hat{V}
\sum_{\sigma_1,\ldots,\sigma_N} \exp \big( \beta \mu \hat{\chi}({\bm \sigma})) 
\\&\phantom{=}
\cdot \int_{\hat{V}^N} \exp  \big( - \beta \hat{E}(\bm{\sigma}; \bm{x}; \hat{V}) \big) d\bm{x} \Bigg).
\end{split}
\end{align}
Our main assumption is that for small temperatures the atoms are near their equilibrium positions, possibly only forming noninteracting defects.

\subsection*{Step 1: Integrating over the volume}
In \eqref{eq:ap1:phi} we fix $\bm{\sigma}$ and assume that $\bm{x}$ are near their equilibrium positions and $V$ is near the equilibrium volume $V_0$.
We then make a change of variables $x_i = V/V_0 \tilde{x}_i$ and $\tilde{E}(\bm{\sigma}; \tilde{\bm{x}}; \hat{V}) = \hat{E}(\bm{\sigma}; \bm{x}; V)$, so that $\tilde{x}_i$ do not change under volumetric expansion.
Then \eqref{eq:ap1:phi} turns into
\begin{align*}
\hat{\Phi} &= -T \log \Bigg( \int_{0}^{\infty} \exp(-\beta p \hat{V}) {\rm d}\hat{V}
\sum_{\sigma_1,\ldots,\sigma_N} \exp \big( \beta \mu \hat{\chi}({\bm \sigma}))
\\&\phantom{=}
\cdot \int_{\hat{V}_0^N} \exp \big( - \beta \tilde{E}(\bm{\sigma}; \tilde{\bm{x}}; \hat{V}) \big) (V/V_0)^N d\tilde{\bm{x}} \Bigg).
\end{align*}
Formally, we cannot set $p=0$ in this integral directly as the integral $\int_{0}^{\infty} \exp(-\beta p \hat{V}) {\rm d}\hat{V}$ would then diverge.
In \eqref{eq:ap1:phi} we fix $\bm{\sigma}$ and $\tilde{\bm{x}}$ and let
\begin{align*}
\hat{\Phi}_{\bm{\sigma},\tilde{\bm{x}}} &=
-T \log \Bigg( \int_{0}^{\infty} \exp(-\beta p \hat{V})
\exp \big( - \beta \tilde{E}(\bm{\sigma}; \tilde{\bm{x}}; \hat{V}) \big)
\\&\,\phantom{\displaystyle=-T \log \int_{0}^{\infty} \mathstrut}\,
\cdot 
(V/V_0)^N 
{\rm d}\hat{V} \Bigg)
,
\end{align*}

so that \eqref{eq:ap1:phi} reads
\begin{equation}\label{eq:ap1:phi_2}
\hat{\Phi} = -T \log
\sum_{\sigma_1,\ldots,\sigma_N}
\exp \big( \beta \mu \hat{\chi}({\bm \sigma}))
\int_{\hat{V}_0^N}
\exp \big(-\beta \hat{\Phi}_{\bm{\sigma},\tilde{\bm{x}}} \big) {\rm d}\tilde{\bm{x}}.
\end{equation}

To expand \eqref{eq:ap1:phi_2}, we use a general formula for the expansion of a Gaussian-like function:
\[
-T \log \int \exp\big( -\beta f(y) + g(y) \big) {\rm d}y.
\]
The expansion is based on the Taylor expansion of $f$ and $g$ around the maximum point of $f$.
Namely, let $y_0 = {\rm argmax}_y\, f(y)$. Then
\begin{align*}
	&\vspace{10em}\textstyle
	-T \log \int \exp\big( -\beta f(y) + g(y) \big) {\rm d}y
	\\ &~~~~~~~~\textstyle
    = f(y_0) + {\textstyle \frac{1}{2}} T \log \left(\beta f''(y_0) / \pi \right) - T g(y_0) + O(T^2).
\end{align*}
To expand the integral in $\hat{\Phi}_{\bm{\sigma},\tilde{\bm{x}}}$, we use the above formula with
$y=\hat{V}$, $f(\hat{V}) = \tilde{E}(\bm{\sigma}; \tilde{\bm{x}}; \hat{V}) + p \hat{V}$ and $g(V) = N \log\Big(\frac{V}{V_0}\Big)$:
\begin{align*}
\hat{\Phi}_{\bm{\sigma},\tilde{\bm{x}}} &=
\tilde{E}(\bm{\sigma}; \tilde{\bm{x}}; \hat{V}_0)
+ p \hat{V_0}
\\&\phantom{=}
+{\textstyle\frac{1}{2}}
T
\log \bigg(
	\beta \frac{{\rm d}^2}{{\rm d}\hat{V}^2} \big(\tilde{E}(\bm{\sigma}; \tilde{\bm{x}}; \hat{V}) + p \hat{V} \big)
\bigg)
\\&\phantom{=}
- {\textstyle\frac{1}{2}} T \log(2\pi)
+ O(T^2)
.
\end{align*}
In what follows, we neglect the $p \hat{V}_0$ term.
Because the left-hand side scales as the first power of $N$ which can be large, we further track explicitly the dependence on $N$,
in which case even the first-order term can be neglected:
\[
\hat{\Phi}_{\bm{\sigma},\tilde{\bm{x}}} =
\tilde{E}(\bm{\sigma}; \tilde{\bm{x}}; \hat{V}_0) 
+ O(T) o(N).
\]
This formula expresses the simple fact that in the free energy of the structure at low temperature we only need its energy at the equilibrium volume.

We recall that $\tilde{x} = x$ at $V=V_0$, and we hence simply have
\begin{align*}
\hat{\Phi} &= - T \log\Bigg(
\sum_{\sigma_1,\ldots,\sigma_N}
\exp \big( \beta \mu \hat{\chi}({\bm \sigma}) \big)
\\&\,\phantom{=- T \log\Bigg(}\,
\cdot \int_{\hat{V}_0^N}
\exp \big(-\beta \hat{E}(\bm{\sigma};\bm{x};\hat{V}_0) \big) {\rm d}\bm{x}\bigg)
\\&\phantom{=\mathstrut}
+ O(T) o(N)
.
\end{align*}
%In the remainder of Appendix \ref{app:free_en_asym} we will drop the dependence of $\hat{E}$ on $\hat{V}_0$.

\subsection*{Step 2: Integrating over atomic positions}
We omit the explicit dependence of $E$ over $\hat{V_0}$, thus writing $\hat{E}(\bm{\sigma};\bm{x};\hat{V}_0) = \hat{E}(\bm{\sigma};\bm{x})$. We separate the atoms into two groups: the last atom $(\sigma_N, x_N)$ and the rest of the atoms with coordinates $\tilde{x}$ (we reuse the tilde superscript for the new, different from the previous subsection, variables $\tilde{x}$). Because of translation symmetry, we can carry out the integral over $x_N$ first and get
\begin{align*}
\hat{\Phi} &= 
- T \log V_0
- T \log\Bigg(
\sum_{\sigma_1,\ldots,\sigma_N}
\exp \big( \beta \mu \hat{\chi}({\bm \sigma}) \big)
\\&\,\phantom{=T \log\Bigg(\mathstrut}\,
\cdot \int_{\hat{V}_0^{N-1}}
\exp \big(-\beta \hat{E}(\bm{\sigma}; \tilde{\bm{x}}; 0) \big) {\rm d}\tilde{\bm{x}}\Bigg)
\\&\phantom{=} + O(T) o(N)
.
\end{align*}
Note that the term $T \log V_0$ can be adsorbed into $O(T) o(N)$.
We fixed the position of the last atom to be $0$ in the integral, which fixes positions of all other atoms to be close to the positions of the ground-state lattice given by coordinates $\xi_1,\ldots,\xi_{N-1}$.
Taking into account that this is one of the $(N-1)!$ possible assignments of atoms to their lattice sites, we have
\begin{align*}
\hat{\Phi} &=  - T \log\big((N-1)!\big)
\\&\phantom{=\mathstrut}
- T \log \Bigg( \sum_{\sigma_1,\ldots,\sigma_N} \exp \big( \beta \mu \hat{\chi}({\bm \sigma}))
\\&\phantom{=-T\mathstrut}\cdot
\int_{x_1\sim\xi_1} \cdots \int_{x_{N-1}\sim\xi_{N-1}} \exp  \big( - \beta \hat{E}(\bm{\sigma}; \tilde{\bm{x}}; 0) \big) {\rm d}\tilde{\bm{x}} \Bigg)  
\\&\phantom{=\mathstrut}
+ O(T) o(N)
,
\end{align*}
where integrating over $\xi_i\sim x_i$ means integrating over some small region near $x_i$.
We note that the equilibrium positions of the structure $\xi_1,\ldots,\xi_{N-1}$ depend on the atomic types $\bm{\sigma}$.
We use that $\log\big((N-1)!\big) = N \log N - N + o(N)$.

We expand the integrals in $x_i$ as in \cite{ladygin2021-phad}:
\begin{widetext}
\begin{align*}
    \hat{\Phi} &=  - T (N \log N - N)
    \\&\phantom{=\mathstrut}
    - T \log \Bigg( \sum_{\sigma_1,\ldots,\sigma_N} \exp \big( \beta \mu \hat{\chi}({\bm \sigma})\big)
    \exp\big(-\beta\hat{E}(\bm{\sigma})\big)
    (2\pi T)^{(3N-3)/2} (\det \hat{H}(\bm{\sigma}))^{-1/2}\Bigg) + O(T) o(N)
    \\&=
    - T (N \log N - N) - {\textstyle \frac32} T N \log(2\pi T)
\\&\phantom{=\mathstrut}
- T \log \Bigg( \sum_{\sigma_1,\ldots,\sigma_N} \exp \big( -\beta \hat{E}(\bm{\sigma}) + \beta \mu \hat{\chi}({\bm \sigma})\big) (\det \hat{H}(\bm{\sigma}))^{-1/2} \Bigg)
+ O(T^2 N) + O(T) o(N)
,
\end{align*}
\end{widetext}
where $\hat{E}(\bm{\sigma})$ is the energy of a configuration given by atomic types $\bm{\sigma}$ whose positions are close to $\xi_1,\ldots,\xi_{N-1},0$ and $\hat{H}(\bm{\sigma})$ is the energy Hessian at the equilibrium positions.

\subsection*{Step 3: Summing over atomic types}

We are interested to find the asymptotic expansion of the free energy near the ideal unary structure with $c=0$ ($c=1$ is treated similarly). 
In this structure the type-2 atoms constitute a substitutional defect whose energy we denote as $E_{\rm def}$.
Then $\hat{E}(\sigma) = N E_0 + N \chi(\sigma) E_{\rm def} + N O(\chi(\sigma)^2)$.
This formula expresses the fact that in the dilute limit $\hat{\chi}(\sigma)\to 0$ the main contributions to the energy are noninteracting defects, and their interaction can be neglected and adsorbed into $O(N \chi(\sigma)^2)$.
We hence have
\begin{align*}
    \hat{\Phi} &= 
    - T (N \log N - N) - {\textstyle \frac32} T N \log(2\pi T)
    \\&\phantom{=\mathstrut}
    - T \log \Bigg( \sum_{\sigma_1,\ldots,\sigma_N}
        \exp \big( -N \beta E_0 \big)
    \\&\phantom{=-T \mathstrut}
        \cdot \exp \big( N \beta (\mu - E_{\rm def}) \chi({\bm \sigma}) + O(N \chi({\bm \sigma})^2) \big)
    \\&\phantom{=-T \mathstrut}
        \cdot (\det \hat{H}(\bm{\sigma}))^{-1/2} \Bigg)
    \\&\phantom{=\mathstrut} 
    + O(T^2 N) + O(T) o(N)
    .
\end{align*}
We postulate that, following the expansion of energy, that the Hessian is expanded in the same way:
$N^{-1}\det \hat{H}(\sigma) = N^{-1} \det \hat{H}_0 + O(\chi(\sigma))$, where $H_0$ is the energy Hessian at the ground state.
Physically this expresses the fact that a defect may only affect the vibrational states locally, affecting the vibrational free energy only in the first order in the concentration of such defects.
The formula is then simplified to
\begin{align}\label{eq:nne_phi_1}
\begin{split}
	\hat{\Phi} &= 
    - T (N \log N - N) - {\textstyle \frac32} T N \log(2\pi T)
    + N E_0
	\\&\phantom{=\mathstrut}
    + {\textstyle \frac12} T \log \det \hat{H}_0
    + \hat{\Phi}_{\sigma} + O(T^2 N) + O(T) o(N),
\end{split}
\end{align}
where $\hat{\Phi}_{\sigma}$ denotes a configurational contribution to the semi-grand potential and is given by
\begin{align*}
    \hat{\Phi}_{\bm \sigma} &= - T \log \Bigg( \sum_{\sigma_1,\ldots,\sigma_N}
    \exp \big( N \beta (\mu - E_{\rm def}) \chi({\bm \sigma})\big)
    \\ &\,\hphantom{\displaystyle= - T \log \sum_{\sigma_1,\ldots,\sigma_N}}\,
     \cdot\exp\big(O(N \chi({\bm \sigma})^2) \big) \Bigg).
\end{align*}
We note that instead of summing over $\sigma_1,\ldots,\sigma_N$ we can sum over the number of type-2 atoms which we denote by $k$:
%\begin{align*}
%\hat{\Phi}_{\bm \sigma} = 
%- T \log \sum_{k=0}^{N} \left( \sum_{\sigma_1,\ldots,\sigma_N \in {\mathcal S}_k }
%\exp \big( -\beta \tilde{E}_{\rm def} k + O(N^{-1} k^2) \big) \right),
%\end{align*}
\begin{align*}
	\hat{\Phi}_{\bm \sigma} =&
	- T \log \sum_{k=0}^{N} \hat{\phi}_{\bm \sigma}(k),
	\qquad\text{where}
	\\
	\hat{\phi}_{\bm \sigma}(k) =& \sum_{\sigma_1,\ldots,\sigma_N \in {\mathcal S}_k }
\exp \big( -\beta \tilde{E}_{\rm def} k + O(N^{-1} k^2) \big).
\end{align*}
Here by ${\mathcal S}_k$ we denote the set of all the configuration with $k$ type-2 atoms and $\tilde{E}_{\rm def} = E_{\rm def} - \mu$.
We assume that $\tilde{E}_{\rm def} > 0$ (otherwise the ideal unary crystal structure would not be stable for that $\mu$).
 After expansion, $\hat{\Phi}_{\bm \sigma}$ has the form
\begin{align}\label{eq:phi_sigma_11}
\hat{\Phi}_{\bm \sigma} =
- T \log \sum_{k=0}^{N} \begin{pmatrix} N \\k \end{pmatrix} 
\exp \big( -\beta \tilde{E}_{\rm def} k + O(N^{-1} k^2) \big),
\end{align}
where we account for the number of configurations with $k$ type-2 atoms using the binomial coefficient $\begin{pmatrix} N \\k \end{pmatrix}$.

We use the Stirling's formula to express
\begin{align*}
\log\begin{pmatrix}N\\k\end{pmatrix}
&= 
N \Big( -k/N \log(k/N) 
\\&\phantom{=\Big(N\mathstrut}
-(N-k)/N \log((N-k)/N)\Big) 
\\&\phantom{=}
+ O(|\log N| + |\log k|)
\end{align*}
We further define $x_k := k/N$ and hence
\begin{align*}
\hat{\Phi}_{\bm \sigma}
&=
- T \log \Bigg( \sum_{k=0}^{N}
	\exp \Big( N \big(-x_k\log(x_k)  
    \\&\phantom{=-T \log T \sum_{k=0}^{N}}
    - (1-x_k)\log(1-x_k) - \beta \tilde{E}_{\rm def} x_k\big) \Big)
    \\&\phantom{=-T \log T T } 
    \cdot \exp\big(O(\log N + |\log x_k|) + O(N x_k^2) \big) \Bigg)
\\&=
- T \log \sum_{k=0}^{N} \exp(f_N(x_k)),
\end{align*}
where we define
\begin{align}\label{eq:f_N}
\begin{split}
f_N(x) &:= - N \big(x\log(x) + (1-x) \log(1-x) + \beta \tilde{E}_{\rm def} x \big) 
\\&\phantom{:=\mathstrut}+ O(\log N + |\log x|) + O(N T^2).
\end{split}
\end{align}
In \eqref{eq:f_N} we use that the squared concentration of the defects is exponentially small in temperature and can be adsorbed into $O(N T^2)$.  
In order to compute the sum analytically, we approximate $f_N(x)$ with its Taylor series up to the second term, in the form $f_N(x) \approx -a (x - b)^2 + q$ near $b = {\rm argmin}_x\, f_N(x)$.  
The equilibrium concentration of the defects is given by $b$ as its represents the most probable state of the system
\begin{align}\label{eq:equil_conc}
    \<\chi\> = b = \frac{1}{1 + \exp \big(\beta \tilde{E}_{\rm def}\big)} + O(N^{-1}), 
\end{align}
while the other coefficients are given by
\begin{align*}
	q &= N \log\left(1 + e^{-\beta \tilde{E}_{\rm def}}\right)
	+ O\left(\log N\right) + O(N T^2), \\
	a &= N \big( 1 + \cosh(\beta \tilde{E}_{\rm def}) \big) + O(1),
\end{align*}
where $O(1)$ denotes a term that does not grow as $N \rightarrow \infty$. 
We next make use of the following quadrature formula
\begin{equation*}\label{eq:sum-to-integral}
N^{-1} \sum_{k=0}^N F(x_k) = \int_0^1 F(x) {\rm d}x + O\big(N^{-1}\big)
\end{equation*}
with $F(x) = \exp(f(x))$.
We can hence calculate
\begin{align}\label{eq:phi_sigma}
\begin{split}
    \hat{\Phi}_{\bm \sigma} &= 
    -T \log \bigg(\sum_{k=0}^{N} \exp(f_N(x_k)) \bigg)
    \\&= 
    -T \log \left(N \int_0^1 \exp(f_N(x_k)) + O(1)\right)
    \\&\approx
     -T \log \left(N \int_0^1 \exp\big(-a (x - b)^2 + q\big) {\rm d}x  + O(1) \right) \\&
    = -T \log \left( N e^q \sqrt{\pi/a} + O(1) \right)\\&
    = -T N \log\big(1 + e^{-\beta \tilde{E}_{\rm def}}\big) + O(T \log N) + O(N T^2).
\end{split}
\end{align}
Using $\hat{\Phi}_{\sigma}$ given by \eqref{eq:phi_sigma} and back-substituting it to \eqref{eq:nne_phi_1}, we obtain
\begin{align*}
\begin{split}
    \hat{\Phi} &= 
    - T (N \log N - N) - {\textstyle \frac32} T N \log(2\pi T)
    + N E_0
    \\&\phantom{=\mathstrut}
    + {\textstyle \frac12} T \log \det \hat{H}_0
    -T N \log\big(1 + e^{- \beta \tilde{E}_{\rm def}}\big)
    \\&\phantom{=\mathstrut}
     + O(T)o(N) + O(N T^2)
    .
\end{split}
\end{align*}

We next perform the Legendre transformation given by \eqref{eq:F} to arrive at the final expression of the crystalline free energy: 
\begin{align}\label{eq:f_final}
\begin{split}
    G &= \Phi + \mu \< \chi \>
    % - p \<V\>
    \\&
       = E_0 (\<\chi\>) + E_{\rm def} \<\chi\>
       - T \log(N)
       + T - {\textstyle \frac32} T \log(2\pi T)
       \\&\phantom{=}
       + T \<\chi\> \log \<\chi\>
       + T (1-\<\chi\>) \log\left(1-\<\chi\>\right)
       \\&\phantom{=}
       + {\textstyle \frac12} T N^{-1} \log\det \hat{H}(\<\chi\>) 
       + O\big(T\big)o(1) + O(T^2), 
\end{split}
\end{align}
where $o(1)$ denotes a vanishing as $N\to\infty$ term. In \eqref{eq:f_final} we use supplementary expressions 
\begin{align*}
	\log \big(1 + e^{ -\beta \tilde{E}_{\rm def}}\big) &= - \log(1 - \<\chi\>), \qquad{\text{and}}
    \\
	\mu &= E_{\rm def} - \beta^{-1} \log \left( \frac{1-\<\chi\>}{\<\chi\>} \right),
\end{align*}
which follow from \eqref{eq:equil_conc}.

\section{Free energy asymptotic of ${\rm Na}_2{\rm K}$}
\label{app:asym_im}

Here we derive the asymptotic free energy for ${\rm Na}_2{\rm K}$ intermetallic alloy, where we denote the concentration of Na atoms as $c_{\rm Na}$. The elementary cell of ${\rm Na}_2{\rm K}$ consists of 12 atoms, although there are only three different substitutional defects possible: K substitutes Na with the formation energies $E_{1}$ (6 sites) and $E_{2}$ (2 sites), while Na substitutes K with the energy $E_3$ (4 sites). We denote the fraction of defect sites to the total number of sites in the elementary cell as $p_1 = \frac12$, $p_2 = \frac16$, and $p_3 = \frac13$. We assume that the defects do not interact with each other as their concentrations are small.

Step 1 and step 2 are the same for intermetallic alloy as in Appendix \ref{app:free_en_asym}. We now focus on step 3 and consider the configurational contribution to the semi-grand potential that is given by
\begin{align}\label{eq:phi_sigma_kna2}
\begin{split}
\hat{\Phi}_{\sigma} &= 
    -{\textstyle \frac23} \mu - T \log
    \sum_{k_1 = 0}^{p_1 N} 
    \sum_{k_2 = 0}^{p_2 N} 
    \sum_{k_3 = 0}^{p_3 N} 
    \binom{p_1 N}{k_1} 
    \binom{p_2 N}{k_2} 
    \binom{p_3 N}{k_3} 
    \\&\phantom{=-{\textstyle \frac23} \mu - T \log}
    \cdot \exp \big( \beta (E_1 + \mu) k_1 + O \big(N^{-1} k_1^2\big) \big)
    \\&\phantom{=-{\textstyle \frac23} \mu - T \log}
    \cdot \exp \big( \beta (E_2 + \mu) k_2 + O \big(N^{-1} k_2^2\big) \big)
    \\&\phantom{=-{\textstyle \frac23} \mu - T \log}
    \cdot \exp \big( \beta (E_3 - \mu) k_3 + O \big(N^{-1} k_3^2\big) \big)
    ,
\end{split}
\end{align}
where $k_i$ denotes the number of the type-$i$ defects and  $\begin{pmatrix} p_i N \\k_i \end{pmatrix}$ accounts for the number of configurations with $k_i$ defects. We rewrite \eqref{eq:phi_sigma_kna2} similarly to \eqref{eq:phi_sigma_11}:   
\begin{align*}
    \hat{\Phi}_{\sigma} &=
    -{\textstyle \frac23} \mu 
    - \sum_{i=1}^{3} T \log \big( \psi_{\sigma}^{(i)} \big), \qquad\text{where}
    \\
    \psi_{\sigma}^{(i)} &= \sum_{k_i = 0}^{p_i N} \binom{p_i N}{k_i} \exp \big(-\beta \tilde{E_i} k_i + O \big( N^{-1} k_i^2 \big) \big) 
    ,
\end{align*}
Here we denote $\tilde{E_1} := E_1 + \mu$, $\tilde{E_2} := E_2 + \mu$, and $\tilde{E_3} := E_3 - \mu$.
We follow step 3 in Appendix \ref{app:free_en_asym} to sum over $k_i$ and arrive at the equation for the equilibrium concentrations of the type-$i$ defects:
\begin{align*}
c_i = c_i(\mu) =  p_i \, \frac{1}{\big( 1 + \exp(\beta \tilde{E_i}) \big)} + O(N^{-1}),
\end{align*}
from which we find $c_{\rm Na} = \frac23 - c_1 - c_2 + c_3$ and the semi-grand thermodynamic potential (or, more precisely, its configurational part):
%\begin{align*}
%	\hat{\Phi}_{\sigma} =& -{\textstyle \frac23} \mu -T\, p_1 \log\left( 1 + \exp(-\beta \tilde{E_1}) \right)  
%	-T \, p_2 \log\left( 1 + \exp(-\beta \tilde{E_2}) \right) \\&
%	-T \, p_3 \log\left( 1 + \exp(-\beta \tilde{E_3}) \right) 
%	+ O(T \log N) + O(N T^2). 
%\end{align*}
\begin{align*}
    \hat{\Phi}_{\sigma}(\mu, T) =&
    -T N \, \sum_{i=1}^3 p_i \log\left( 1 + \exp \big(-\beta \tilde{E}_i \big) \right)  
    \\&
    -{\textstyle \frac23} N \mu + O(T \log N) + O(N T^2). 
\end{align*}

Our next step is to find $\mu = \mu(c_{\rm Na},T)$ as the minimizer of $\hat{\Phi}_{\sigma}(\mu, T)$ subject to the constraint $\frac23 - c_1(\mu) - c_2(\mu) + c_3(\mu) = c_{\rm Na}$ and thus obtain the configurational contribution to the free energy
\[
G_{\sigma}(c_{\rm Na}, T) = \Phi_{\sigma}(\mu(c_{\rm Na},T), T) - \mu c_{\rm Na},
\]
and subsequently the full free energy (G) by doing similar steps as in Appendix \ref{app:free_en_asym}
\begin{align*}
G &=  E_0 - {\textstyle \frac32} T \log(2\pi T) + {\textstyle \frac12} T N^{-1} \log\det \hat{H}_0 + F_{\sigma} 
\\&\phantom{=}
+ O(T^2) + O(T)o(1),
\end{align*}
where where $o(1)$ denotes a vanishing as $N\to\infty$ term. 

We do these manipulations in the Mathematica symbolic algebra software \cite{mathematica} and obtain the following formula:
\begin{align*}
	&\vspace{-1em}\textstyle
	G^{\rm im} =
	E_0 - {\textstyle \frac32} T \log(2\pi T) + {\textstyle \frac12} T N^{-1} \log\det \hat{H}_0
	\\ &~\textstyle
	- \frac{\sqrt{2}}{3} \sqrt{e^{\beta  E_1}+3 e^{\beta  E_2}} e^{-\frac{1}{2} \beta  (E_1 + E_2 + E_3)}
	\\ &~\textstyle
	+ 
	\frac{1}{2}  \left(\log \left(\frac{e^{\beta  E_1}+3 e^{\beta  E_2} }{2} \right)-\beta  (E_1 + E_2 - E_3)\right) \left(c_{\rm Na} - \frac{2}{3}\right)
	\\&~\textstyle
	+ 
	\frac{3}{2 \sqrt{2}} \frac{e^{\frac{1}{2} \beta  (E_1 + E_2 + E_3)}}{\sqrt{e^{\beta E_1}+3 e^{\beta E_2}}} \left(c_{\rm Na} - \frac{2}{3}\right)^2 
    + O\Big(\big(c_{\rm Na} - {\textstyle \frac{2}{3}}\big)^3\Big) 
    \\&~\textstyle
    + O(T^2) + O(T)o(1)
    .
\end{align*} 

\section{Derivatives of the free energy}
\label{appd:entrop_deriv}

\subsection{Derivative with respect to concentration}

We start by differentiating \eqref{eq:F} with respect to $\mu$:
\begin{align}\label{eq:der_wrt_c_1}
	\frac{\partial \hat{G}}{\partial \<\chi\>} \frac{\partial \<\chi\>_{T,\mu}}{\partial \mu}
	=
	\frac{\partial \hat{\Phi}(T,\mu)}{\partial \mu}
	+ \<\hat{\chi}\>_{T,\mu}
	+ \mu \frac{\partial \<\hat{\chi}\>_{T,\mu}}{\partial \mu}.
\end{align}
We next calculate the first term on the right-hand side:
\begin{align*}
	\frac{\partial \hat{\Phi}(T,\mu)}{\partial \mu}
	&=
	-\beta^{-1} \frac{ \partial \log(\hat{Z})}{\partial \mu}
	\\ &=
	-\beta^{-1} \hat{Z}^{-1} \frac{ \partial \hat{Z}}{\partial \mu}
	\\ &=
	-\beta^{-1} \hat{Z}^{-1} \sumint_{\cfg} \exp\big(-\beta (\hat{E}(\cfg) - \mu \hat{\chi}(\cfg)) \big) 
    \\ &\phantom{=-\beta^{-1} \hat{Z}^{-1} Z^{-1}}
    \cdot \frac{ \partial (\beta \mu \hat{\chi}(\cfg))}{\partial \mu}
	\\ &=
    - \< \hat{\chi} \>_{T,\mu}.
\end{align*}
Hence \eqref{eq:der_wrt_c_1} transforms to 
\[
\frac{\partial \hat{G}}{\partial \<\hat{\chi}\>} \frac{\partial \<\hat{\chi}\>_{T,\mu}}{\partial \mu}
=
\mu \frac{\partial \<\hat{\chi}\>_{T,\mu}}{\partial \mu}.
\]
We exclude the unphysical case in which a change in $\mu$ does not induce a change in concentration, therefore we have that $\frac{\partial \<\hat{\chi}\>_{T,\mu}}{\partial \mu} \ne 0$. 
We hence arrive to
\begin{align}\label{eq:der_f_c}
    \frac{\partial G}{\partial \<\chi\>} = \mu.     
\end{align}

\subsection{Derivative with respect to temperature}

We start by deriving the expression for ${\partial \big( \beta \hat{\Phi}(T,\mu) \big)}/{\partial \beta}$:
\begin{align*}
	\frac{\partial \big( \beta \hat{\Phi}(T, \mu) \big)}{\partial \beta}
	&=
	- \frac{\partial \log \hat{Z}}{\partial \beta} 
	\\ &=
	- Z^{-1} \frac{\partial \hat{Z}}{\partial \beta}
	\\ &=
	Z^{-1} \sumint_{\cfg} (\hat{E}(\cfg) - \mu \hat{\chi}(\cfg)) 
    \\ &\phantom{=Z^{-1} \sumint_{\cfg}}
    \cdot \exp \big(-\beta (\hat{E}(\cfg) - \mu \hat{\chi}(\cfg)) \big) 
	\\ &=
	\<\hat{E}\>_{T,\mu} - \mu \<\hat{\chi}\>_{T,\mu},
\end{align*}
hence
\begin{align}\label{eq:der_phi_beta}
\frac{\partial \big( \beta \hat{\Phi}(T, \mu) \big)}{\partial \beta} = 
\<\hat{E}\>_{T,\mu} - \mu \<\hat{\chi}\>_{T,\mu}.
\end{align}
Let us now multiply both sides of \eqref{eq:F} by $\beta$ and differentiate resulting equation with respect to $\beta$:
\[
\beta \hat{G}(T, \<\chi\>_{T,\mu}) = \beta \hat{\Phi}(T, \mu) + \beta \big( \mu \<\hat{\chi}\>_{T,\mu} \big),
\]
\[
\frac{\partial \big(\beta \hat{G}\big)}{\partial \beta} + \beta \frac{\partial \hat{G}}{\<\hat{\chi}\>} \frac{{\partial \<\hat{\chi}\>_{T,\mu}}}{\partial \beta} = 
\frac{\partial \big( \beta \hat{\Phi}\big)}{\partial \beta} + \mu \<\hat{\chi}\>_{T,\mu} + \beta \mu \frac{{\partial \<\hat{\chi}\>_{T,\mu}}}{\partial \beta}.
\]
We use \eqref{eq:der_f_c} and \eqref{eq:der_phi_beta} and arrive to  
\begin{align*}
\frac{\partial \big(\beta G \big)}{\partial \beta}
= \< E \>_{T, \mu}.
\end{align*}
We next transition from differentiating with respect to $\beta$ to differentiating with respect to $T$, which leads to
\begin{align}\label{eq:der_f_T}
    \frac{\partial \big(\beta G \big)}{\partial T}
    = - \frac{\< E \>_{T, \mu}}{T^2}.
\end{align}

\section{Uncertainty in the input data}
\label{appd:error_input_data}

In the simulation we find the averaged values, $\overline{E}$ and $\overline{c}$, however, the data given to the Gaussian process, according to \eqref{eq:der_f_c} and \eqref{eq:der_f_T} is $\overline{E}$ and $\mu$.
Here we derive how to convert the uncertainty in $\overline{E}$ and $\overline{c}$ into the uncertainties for $\overline{E}$ and $\mu$.

We obtain the trajectory-averaged mean energy $\overline{E}$ and concentration $\overline{c}$ from MD simulations in the $\mu$pT ensemble with uncertainties. The mean energy and concentration form the joint Gaussian distribution 
\begin{align*}\label{eq:e_c_gaus_distr}
\begin{pmatrix}
    \,\overline{E}\, \\
    \overline{c}
\end{pmatrix} 
&\sim \mathcal{N}
\begin{pmatrix}
    \begin{pmatrix}
    E_0 \\
    c_0
    \end{pmatrix}
    ,
    \begin{pmatrix}
    {\rm cov}(\overline{E}, \overline{E}) & {\rm cov}(\overline{E}, \overline{c}) \\
    {\rm cov}(\overline{E}, \overline{c}) & {\rm cov}(\overline{c}, \overline{c})
    \end{pmatrix}
\end{pmatrix}.
\end{align*}
Since we condition our algorithm based on Gaussian process to satisfy the equations \eqref{eq:der_f_c} and \eqref{eq:der_f_T}, we need to convert the uncertainty in $\overline{E}$ and concentration $\overline{c}$ into the uncertainty in \eqref{eq:der_f_c} and \eqref{eq:der_f_T}, in other words, determine the uncertainty in
\[
{\bm q} = 
\begin{pmatrix}
	\dfrac{\partial (\beta G)}{\partial T\vphantom{\big|}} (\overline{c}, T) + \dfrac{\overline{E}}{T^2}
\\
    \dfrac{\partial G \vphantom{\big|}}{\partial c} (\overline{c}, T) - \mu 
\end{pmatrix}.
\]
To that end, we expand the derivatives of $G$ around $(c_0, T)$:
\[
{\bm q} \approx 
\begin{pmatrix}
 	\dfrac{\partial (\beta G)}{\partial T \vphantom{\big|}} (c_0, T)
	+ \dfrac{\partial^2 (\beta G)}{\partial T \partial c\vphantom{\big|}} (c_0, T)\,(\overline{c}-c_0) + \dfrac{\overline{E}}{T^2}
	\\
	\dfrac{\partial G\vphantom{\big|}}{\partial c} (c_0, T)
	+ \dfrac{\partial^2 G\vphantom{\big|}}{\partial c^2} (c_0, T) (c_0, T) \,(\overline{c}-c_0) - \mu.
\end{pmatrix},
\]
We next use that $\frac{\partial (\beta G)}{\partial T} (c_0, T) = -\frac{E_0}{T^2}$ and $\frac{\partial G}{\partial c} (c_0, T) = \mu$:
\[
{\bm q} \approx
\begin{pmatrix}
	\dfrac{\overline{E} - E_0}{T^2}
	+ \dfrac{\partial^2 (\beta G)}{\partial T \partial c\vphantom{\big|}}\,(\overline{c}-c_0)
	\\
	\dfrac{\partial^2 G\vphantom{\big|}}{\partial c^2} (c_0, T)\,(\overline{c}-c_0)
\end{pmatrix},
\]
or in the matrix form
\begin{equation*}\label{eq:linear_trans_A}
{\bm q} \approx
    A
    \begin{pmatrix}
        \,\overline{E} - E_0 \\
        \overline{c} - c_0
    \end{pmatrix},
\end{equation*}
where 
\[
A = 
\begin{pmatrix}
    \dfrac{1}{T^2} & \dfrac{\partial^2 (\beta G)}{\partial T \partial c \vphantom{\big|}} \\
    0 & \dfrac{\partial^2 G \vphantom{\big|}}{\partial c^2}
\end{pmatrix}.
\]

We apply the linear transformation to the Gaussian distribution $\begin{pmatrix}
    \,\overline{E} - E_0 \\
    \overline{c} - c_0
\end{pmatrix}$ using the matrix A and obtain the following distribution for ${\bm q}$:
\begin{align*}
{\bm q}  &\sim 
\mathcal{N}
\begin{pmatrix}
    \begin{pmatrix}
    0 \\[10pt]
    0
    \end{pmatrix}
    , 
    \begin{pmatrix}
        k_{11} &  k_{12}\\[10pt]
        k_{21} &  k_{22}
        \end{pmatrix}
\end{pmatrix},
\end{align*}
where the elements of the covariance matrix are given by
\begin{align*}
    k_{11} &= \frac{{\rm cov}\big(\overline{E}, \overline{E}\big)}{T^4}  + \frac{2}{T^2} \dfrac{\partial^2 (\beta G)}{\partial T \partial c} {\rm cov}\big(\overline{E}, \overline{c}\big) 
    \\&\phantom{=\mathstrut}
    +\left( \frac{\partial^2 (\beta G)}{\partial T \partial c}\right)^2  {\rm cov}(\overline{c}, \overline{c}), \\
    k_{12} = k_{21} &= 
    \frac{\partial^2 G}{\partial c^2} \left( 
    \frac{{\rm cov}\big(\overline{E}, \overline{c}\big)}{T^2}  + \frac{\partial^2 (\beta G)}{\partial T \partial c} {\rm cov}(\overline{c}, \overline{c}) \right), \\
    k_{22} &= \left( \dfrac{\partial^2 G}{\partial c^2} \right)^2 {\rm cov}(\overline{c}, \overline{c}).
\end{align*}
The uncertainties of the input data are given by the $k_{11}$ and $k_{22}$ terms of the covariance matrix.

\section{Covariance matrix in equation \eqref{eq:k_J_J}}
\label{appd:err_system_of_lin_eq}
The covariance matrix $K_s = {\rm cov}\big( \<\nabla_S \mathcal{K}, S\>, \<\nabla_S \mathcal{K}, S\> \big)$  from \eqref{eq:k_J_J} has the form
\begin{align*}
    K_s 
    &= 
    \begin{pmatrix}
    k_{11} & k_{12} \\
    k_{21} & k_{22} \\
    \end{pmatrix}
    \\&=
    {\rm cov} 
    \big(\<\nabla_S \mathcal{K}, S\>, \<\nabla_S \mathcal{K},S\> \big)
\end{align*}
We assume that the first phase is solid, with kernel $k_{\rm sol} = k_{\rm sol}\big(T^{(1)}, c^{(1)}; T^{(2)}, c^{(2)}\big)$, and the second phase is liquid, with kernel $k_{\rm liq} = k_{\rm liq}\big( T^{(1)}, c^{(1)}; T^{(2)}, c^{(2)}\big)$.
Substituting $\mathcal{K}_1$ and $\mathcal{K}_2$ from \eqref{eq:sys_eq_1}--\eqref{eq:sys_eq_2} into the above covariance matrix yields the following formulas for the elements of $K_s$:
\begin{align*}
    k_{11} 
    &= \cov \big( \<\nabla_S \mathcal{K}_1, S(c^{(1)})\>, \<\nabla_S \mathcal{K}_1, S(c^{(2)})\> \big) 
    \\&= 
    \frac{\partial}{\partial c^{(1)} \partial c^{(2)}} k_\liq
    +
    \frac{\partial}{\partial c^{(1)} \partial c^{(2)}} k_\sol,
\end{align*}

\begin{align*}
    k_{12} &= {\rm cov} \big( \<\nabla_S \mathcal{K}_1, S(c^{(1)})\>, \<\nabla_S \mathcal{K}_2, S(c^{(2)})\> \big)
    \\&= 
    \cov \bigg(
    -\frac{\der S^{(1)}_\liq}{\der c} + \frac{\der S^{(1)}_\sol}{\der c}
    ,
    \\&\phantom{=\cov (((}
    - S^{(2)}_\liq + S^{(2)}_\sol + c^{(2)}_\liq \frac{\der S^{(2)}_\liq}{\der c} - c^{(2)}_\sol \frac{\der S^{(2)}_\sol}{\der c} 
    \bigg)
    \\&= 
    \frac{\partial}{\partial c^{(1)}} k_\liq
    - 
    c^{(2)}_\liq \frac{\partial^2}{\partial c^{(1)} \partial c^{(2)}} k_\liq 
    \\&\phantom{=\mathstrut} 
    + \frac{\partial}{\partial c^{(1)}} k_\sol
    - c^{(2)}_\sol \frac{\partial^2}{\partial c^{(1)} \partial c^{(2)}} k_\sol, 
\end{align*}
\begin{align*}
    k_{21} &= \cov \big( \<\nabla_S \mathcal{K}_2, S(c^{(1)})\>, \<\nabla_S \mathcal{K}_1, S(c^{(2)})\> \big) = 
    \\&= 
    \cov \bigg(
    - S^{(1)}_\liq + S^{(1)}_\sol + c^{(1)}_\liq \frac{\der S^{(1)}_\liq}{\der c} - c^{(1)}_\sol \frac{\der S^{(1)}_\sol}{\der c},
    \\&\phantom{=\cov (((}
    -\frac{\der S^{(2)}_\liq}{\der c}+ \frac{\der S^{(2)}_\sol}{\der c}
    \bigg)
    \\&= 
    \frac{\partial}{\partial c^{(2)}} k_\liq
    - 
    c^{(1)}_\liq \frac{\partial^2}{\partial c^{(1)} \partial c^{(2)}} k_\liq 
    \\&\phantom{=\mathstrut}
    +
    \frac{\partial}{\partial c^{(2)}} k_\sol
    -
    c^{(1)}_\sol \frac{\partial^2}{\partial c^{(1)} \partial c^{(2)}} k_\sol, 
\end{align*}
\begin{align*}
    k_{22} &=
    \cov \big( \<\nabla_S \mathcal{K}_2, S(c^{(1)})\>, \<\nabla_S \mathcal{K}_2, S(c^{(2)})\> \big) = 
    \\&= 
    \cov \bigg(
    - S^{(1)}_\liq + S^{(1)}_\sol + c^{(1)}_\liq \frac{\der S^{(1)}_\liq}{\der c} - c^{(1)}_\sol \frac{\der S^{(1)}_\sol}{\der c},
    \\&\phantom{=\cov (((}
    - S^{(2)}_\liq + S^{(2)}_\sol + c^{(2)}_\liq \frac{\der S^{(2)}_\liq}{\der c} - c^{(2)}_\sol \frac{\der S^{(2)}_\sol}{\der c} 
    \bigg)
    \\&= 
    k_\liq
    - 
    c^{(2)}_\liq \frac{\partial}{\partial c^{(2)}} k_\liq
    -
    c^{(1)}_\liq \frac{\partial}{\partial c^{(1)}} k_\liq 
    \\&\phantom{=\mathstrut}
    + k_\sol
    - 
    c^{(2)}_\sol \frac{\partial}{\partial c^{(2)}} k_\sol
    -
    c^{(1)}_\sol \frac{\partial}{\partial c^{(1)}} k_\sol
    \\&\phantom{=\mathstrut}
    +
    c^{(1)}_\liq c^{(2)}_\liq \frac{\partial^2}{\partial c^{(1)} \partial c^{(2)}} k_\liq
    +
    c^{(1)}_\sol c^{(2)}_\sol \frac{\partial^2}{\partial c^{(1)} \partial c^{(2)}} k_\sol.
\end{align*}

\bibliography{prb_article.bib}

\end{document}